\documentclass[11pt]{article}

\usepackage{epsf}
\usepackage{color}
\usepackage{bm}
\usepackage{graphicx}
\usepackage{amsmath,amsfonts,amsbsy,amssymb}
\usepackage{indentfirst}
\begin{document}

	\begin{center}   
	{\bf Time-Reversal Symmetry in Non-Hermitian Systems\\}    
\vspace{0.3cm} 	
{Masatoshi Sato$^1$,   Kazuki Hasebe$^2$, 
Kenta Esaki$^1$ and  Mahito Kohmoto$^1$  \\}  
	\vspace{0.2cm}    
{\it \small $^1$  Institute for Solid State Physics, University of Tokyo, Kashiwanoha 5-1-5,  Kashiwa, \\Chiba 277-8581, Japan  }\\  
\vspace{0.2cm}  
	{\it \small  $^2$  
	Department of General Education, Kagawa National College of Technology,    
         Mitoyo, \\Kagawa 769-1192, Japan }   
\vspace{0.2cm}      

\vspace{0.2cm}  
\today 
\end{center}   



\begin{abstract}
For ordinary hermitian Hamiltonians, 
the states show the Kramers
 degeneracy when the system has a half-odd-integer spin and the
time reversal  operator obeys $\Theta^2=-1$, but no
such a degeneracy exists when $\Theta^2=+1$.
Here we point out that for non-hermitian systems, there exists  
a degeneracy similar to Kramers even when $\Theta^2=+1$. 
It is found that the new degeneracy follows from the mathematical
 structure of split-quaternion, instead of quaternion from which the
 Kramers degeneracy follows in the usual hermitian cases.
Furthermore, we also show that particle/hole symmetry gives rise to 
a pair of states with opposite energies
on the basis of the split-quaternion in a class of non-hermitian Hamiltonians.
As concrete examples, we examine in detail $N\times N$ 
Hamiltonians with $N=2$ and $4$ which are non-hermitian generalizations of 
spin $1/2$ Hamiltonian and quadrupole Hamiltonian of spin $3/2$, respectively.

\end{abstract}


\tableofcontents

\section{Introduction}\label{sectintro}

The original observation between time-reversal (TR) invariance and
statistical mechanics is traced back to Dyson who pointed out that TR
operator $\Theta$ is naturally incorporated in the algebra of quaternions
 if the system has a half-odd-integer spin and $\Theta^2=-1$ \cite{Dyson1962}. 
He showed that the Kramers degeneracy comes from mathematical structures of quaternion, and its statistical properties are described by the symplectic group.
On the basis of these,  Avron et al. explored topological properties of fermionic systems
with TR symmetry \cite{Avron1988}, and the second Chern number was introduced 
as an extension of the TKNN topological number of quantum Hall effect
\cite{TKNN1982,Kohmoto1985}.   
The TR symmetry and the resultant Kramers degeneracy also play a central role
in recent developments of the quantum spin Hall effects 
\cite{Bernevig2005,Kane2005} and topological insulators
\cite{FKM07, MB07, Roy09}. 
Indeed, the Kramers degeneracy enables us to introduce a new class of
topological numbers characterizing these phases. 
The mathematical structures of the topological insulators have been studied in 
Refs.\cite{Qietal2008,Schnyderetal2008,Kitaev2008}. 

Meanwhile, if the system has an
integer spin and $\Theta^2=+1$ such as boson systems 
and systems with even number of electrons, we have no such a Kramers
degeneracy. 
Correspondingly, its topological structure is rather simple and
the Hamiltonian has a real structure in general.
However, such a consequence changes
if we allow non-hermiticity of
Hamiltonians.
Indeed, as is shown below, there is 
generally a degeneracy similar to the Kramers even when $\Theta^2=+1$ in a class of non-hermitian Hamiltonians. 

Although we usually suppose hermiticity of Hamiltonian,
non-hermitian Hamiltonians also have applications to interesting
problems such as open chaotic
scattering \cite{FS97}, dissipative quantum maps \cite{GT85},
and delocalization of pinned vortices in
superconductors \cite{HatanoNelson199678}. 
We also have non-hermitian Hamiltonians as
effective theories of hermitian systems.
Moreover, non-hermitian Hamiltonians might be meaningful 
themselves if a kind of TR symmetry
such as $\mathcal{PT}$ symmetry \cite{BenderBoettcher1998}
or pseudo-hermiticity \cite{Mostafazadeh2008} is imposed.
They are a part of the motivations that we pursue the present work.  

We investigate TR symmetry with $\Theta^2=+1$ in
non-hermitian Hamiltonians. 
From a general argument, it is shown that such symmetry 
is naturally incorporated in the algebra of
split-quaternion, instead of quaternion.
(See also related work \cite{BG11,BG11_3} in $\mathcal{PT}$ symmetric quantum
mechanics.)
Then a new kind of degeneracy is obtained from
 structures of split-quaternion. 
As concrete examples, we examine $N\times N$ non-hermitian Hamiltonians up
to $N=4$.
The structure of split-quaternion is identified in these Hamiltonians,
and we find that it
has a close similarity to the quaternion structure of the spin $1/2$ Hamiltonian and quadrupole Hamiltonian of spin $3/2$.
Furthermore, it is shown that the particle/hole symmetry also gives rise to
a pair of states with opposite energies $(E,-E)$ in a class of non-hermitian
 Hamiltonians. 
Random matrix classification of the non-hermitian models is also provided.  

The organization of the paper is as follows. In
Sec.\ref{splitquaernionsandTreversal}, a generalized Kramers degeneracy
in pseudo-Hermitian quantum mechanics is discussed. We point out
relations between split-quaternion and TR operation for $\Theta^2=+1$,
and show the existence of generalized Kramers degeneracy in
pseudo-hermitian systems. We also argue relations between the
particle/hole symmetry and split-quaternions.  The split-quaternion
structure of integer spin systems is clarified, too. In
Sec.\ref{sectoinrandom}, we show how the generalized Kramers theorem in
pseudo-hermitian systems is incorporated in the non-hermitian random
matrix classification.  
As a concrete example of pseudo-hermitian model with particle/hole
symmetry, $SU(1,1)$ model is introduced in
Sec.\ref{randommatrixcategorysu11}, and basic properties of the model
are investigated.  
In Sec.\ref{randommatrixcategoryso32}, we argue properties of the
$SO(3,2)$  model with time reversal symmetry $\Theta^2=+1$ as a simple
exemplification of the generalized Kramers degeneracy.  It is also
shown that the $SO(3,2)$ model is realized as a $SU(1,1)$ quadrupole
model with $SU(1,1)$ spin 3/2. Sec.\ref{sectsummary} is devoted to
summary and discussions.

\section{Generalized Kramers degeneracy and split-quaternion} 
\label{splitquaernionsandTreversal}

\subsection{Split-quaternion and time-reversal symmetry}

Let us start with a brief review of the
split-quaternion. 
The split-quaternion \cite{Cockle1848} is a variant of the quaternion
\cite{Hamilton1844} which is written as
\begin{eqnarray}
q=w+xi+yj+zk,
\end{eqnarray}
with real numbers $(w,x,y,z)$ in the basis $(1,i,j,k)$. 
The algebra of the basis for the split-quaternion is
different from that for the quaternion, and it is given by
\begin{eqnarray}
ij=k=-ji,
\quad
jk=-i=-kj,
\quad
ki=j=-ik,
\quad
i^2=-1,
\quad
j^2=1,
\quad
k^2=1. 
\label{eq:ijk}
\end{eqnarray}
Note that $j^2=1$ and $k^2=1$, not $j^2=-1$
and $k^2=-1$ as in the quaternion case.

It has been known that the structure of the quaternion naturally arises
in time-reversal (TR) invariant systems. 
The TR operator is antiunitary, and anticommutes with $i$: 
\begin{eqnarray}
\Theta=U K 
\end{eqnarray}
where $U$ is a unitary operator and $K$ complex conjugates everything to its right. 
For systems with an integer spin, we have
$\Theta^2=+1$, while for systems with a half integer spin, $\Theta^2=-1$.\footnote{ 
The action of the TR operator is two, {\it i.e.} 
the system comes back to the original if we apply $\Theta$ twice.
Thus $\Theta^2$ should be a phase $e^{i\alpha}$,
which implies $U=e^{i\alpha} U^{T}$ and $U^{T}=Ue^{i\alpha}$. 
This yields $U=Ue^{2i\alpha}$, so the phase is $e^{i\alpha}=\pm 1$.}
In the latter case, the TR invariance results in the
structure of quaternion for the Hamiltonian. Then, what is a natural
mathematical structure in the former ?

The answer is the split-quaternion.
The TR operator is antiunitary 
\begin{eqnarray}
\Theta i = -i \Theta. 
\label{eq:thetai}
\end{eqnarray} 
By identifying $j$ and $k$ with $\Theta$ and $i\Theta$,
respectively, one finds a correspondence between the
triplet of the TR algebra and the split-quaternion,
\begin{equation}
(i,\Theta,i\Theta)~\leftrightarrow~ (i,j,k). 
\end{equation} 
Thus, the split-quaternion also fits into 
the TR symmetry with $\Theta^2=+1$. 

In spite of the argument above, it has been
known that there is no such a split-quaternion structure in the
TR invariant Hamiltonians with $\Theta^2=+1$. 
The Hamiltonian supports only a real structure instead \cite{Dyson1962}.
This is because usually the Hamiltonians are supposed to be hermitian.
This implicit assumption makes the split-quaternion into a
real number.
Nevertheless, physical phenomena are not always described by hermitian
Hamiltonians. 
Non-hermitian Hamiltonians also have interesting physical
applications.
Then, if we consider a class of non-hermitian Hamiltonians,  
the hidden split-quaternion structure becomes evident, as will be shown in the following sections.

\subsection{Pseudo-hermiticity}\label{subsecpseudokramers}

A non-hermitian Hamiltonian $H$ is called pseudo-hermitian \cite{Mostafazadeh2008}, when it
satisfies pseudo-hermiticity
\begin{eqnarray}
H^{\dagger}=\eta H \eta^{-1}, 
\label{pseudohermite}
\end{eqnarray}
where $\eta$ is a hermitian operator referred to as the {\em metric operator}.
For example, consider a nonunitary transformation $G$ on 
a hermitian Hamiltonian $H_0$, then  $H=GH_0
G^{-1}$ is not hermitian, but pseudo-hermitian,
\begin{eqnarray}
H^{\dagger}=G^{\dagger -1}H_0 G^{\dagger}=G^{\dagger -1}G^{-1}H
 GG^{\dagger},  
\end{eqnarray}
with a metric operator $\eta=(GG^{\dagger})^{-1}.$

The reason why $\eta$ is called the metric operator is that 
the time-independent inner product of a state is given by a metric $\eta$.
For a non-hermitian Hamiltonian, 
there generally exists a set of states,  $|\phi\rangle$ and
$|\varphi \rangle\!\rangle$ that satisfy \cite{FaisalandMoloney1981}     
\begin{equation}
i\frac{\partial}{\partial t}|\phi\rangle =H|\phi\rangle, ~~~~~ 
i\frac{\partial}{\partial t}|\varphi\rangle\!\rangle
=H^{\dagger}|\varphi\rangle\!\rangle. 
\end{equation}

The time independent inner product is constructed as $\langle \phi|\varphi\rangle\!\rangle$, as shown by  
\begin{equation}
 i\frac{\partial}{\partial t}\langle \phi|\varphi\rangle\!\rangle 
= i\frac{\partial \langle \phi|}{\partial t}|\varphi\rangle\!\rangle
+i\langle \phi|  \frac{\partial  |\varphi\rangle\!\rangle}{\partial t}
=-\langle \phi|H^{\dagger}|\varphi\rangle\!\rangle
+\langle \phi|H^{\dagger}|\varphi\rangle\!\rangle=0.   
\end{equation} 

For a pseudo-hermitian Hamiltonian $H$, (\ref{pseudohermite}) leads to 
\begin{equation}
i\frac{\partial}{\partial t}\eta |\phi\rangle
 =H^{\dagger}\eta|\phi\rangle, ~~~~~ 
i\frac{\partial}{\partial t}\eta^{-1}|\varphi\rangle\!\rangle
=H\eta^{-1}|\varphi\rangle\!\rangle. 
\end{equation}
Thus, we also have additional time-independent inner products, $\langle
\phi|\eta|\phi\rangle$, $\langle\!\langle
\varphi|\eta^{-1}|\varphi\rangle\!\rangle$.
In the following, we mainly use $\langle \phi|\varphi\rangle\!\rangle$
as the inner product unless explicitly written. 

In order for the pseudo-hermiticity to be consistent with the TR
symmetry, the condition (\ref{pseudohermite}) should be commutative with the
TR operation.  
This leads to $\eta^*=\pm U^{\dagger}\eta U$, that is $\Theta \eta =\pm
\eta\Theta$. Therefore, possible metric operators are
classified into two: The first one satisfies $[\eta, \Theta]=0$, and the
second $\{\eta,\Theta\}=0$. 
Here we note that the latter case is proper for  only non-hermitian
Hamiltonians. For hermitian Hamiltonians, we have
$\eta=1$. Thus $\eta$ commutes with $\Theta$ rather trivially. 

\subsection{Generalized Kramers degeneracy}
\label{generalizedKramers}

Let us assume two conditions: one is  the TR symmetry with $\Theta^2=+1$
\begin{equation}
[H,{\Theta}]=0, 
\label{commutativeHThetaB}
\end{equation}
and the other is the anticommutation relation   
\begin{equation}
\{\eta,{\Theta}\}=0.  
\label{anticommutethetaeta}
\end{equation}
Let $|\phi_n\rangle$ 
be eigenstates of $H$,
\begin{eqnarray}
H|\phi_n\rangle=E_n|\phi_n\rangle.
\label{eq:defphin}
\end{eqnarray} 
Then  the corresponding eigenstates $|\phi_n\rangle\!\rangle$ 
of $H^{\dagger}$,
\begin{eqnarray}
H^{\dagger}|\phi_n\rangle\!\rangle=E_n^*|\phi_n\rangle\!\rangle,
\label{eq:defphin2}
\end{eqnarray} 
which satisfy
\begin{eqnarray}
\langle \phi_n|\phi_m\rangle\!\rangle=\langle\!\langle
 \phi_m|\phi_n\rangle=\delta_{nm}. 
\label{eq:bio}
\end{eqnarray}
The eigenstates $|\phi_n\rangle$ and $|\phi_n\rangle\!\rangle$ satisfying
(\ref{eq:bio}) are known as the bi-orthonormal basis {\cite{Wong1967,FaisalandMoloney1981}. 
By using the pseudo-hermiticity (\ref{pseudohermite}), (\ref{eq:defphin2}) is rewritten as 
\begin{equation}
H\eta^{-1}|\phi_n\rangle\!\rangle=E_n^*\eta^{-1}|\phi_n\rangle\!\rangle. 
\label{hetaequation}
\end{equation}
We apply $\Theta$ from the left to both sides of (\ref{hetaequation}) to have 
\begin{equation}
H\Theta\eta^{-1}|\phi_n\rangle\!\rangle=E_n\Theta\eta^{-1}|\phi_n\rangle\!\rangle,  
\end{equation}
where on the left-hand side, we utilized the time reversal invariance of the Hamiltonian (\ref{commutativeHThetaB}). 
Thus $|\phi_n\rangle$ and
$\Theta\eta^{-1}|\phi_n\rangle\!\rangle$ have the same
eigenvalue $E_n$.
Therefore, if they 
are linearly independent, we have degeneracy in the eigenvalues of
 $H$. Note
\begin{equation}
\langle\!\langle \phi_n|\Theta\eta^{-1}|\phi_n\rangle\!\rangle =
\langle\!\langle \Theta^{2}\eta^{-1} 
\phi_n|{\Theta}\phi_n\rangle\!\rangle 
=\langle\!\langle  \phi_n |\eta^{{-1}}
\Theta \phi_n\rangle\!\rangle 
=-\langle\!\langle \phi_n |\Theta\eta^{-1}| \phi_n\rangle\!\rangle. 
\label{eq:Kramerseta}
\end{equation}
In the first equation, we have used the antiunitary property of  
${\Theta}$, and 
the second equation follows from $\Theta^2=+1$ and the hermiticity of $\eta^{-1}$. In the third equation, the anticommutation relation between  ${\Theta}$ and $\eta^{-1}$ was utilized
\footnote{For usual hermitian Hamiltonians, we have
$\eta=1$ in general. Thus, the anticommutativity (\ref{anticommutethetaeta}) does
not hold. This is the reason why  even if a hermitian Hamiltonian has TR
symmetry with $\Theta^2=+1$, there is no generalized Kramers
pair.}.  
Thus we have $\langle\!\langle \phi_n
|\Theta\eta^{-1}| \phi_n\rangle\!\rangle=0$. 
On the other hand,  $\langle\!\langle \phi_n|\phi_n\rangle=1$
from (\ref{eq:bio}).
Therefore $|\phi_n\rangle$ and
$\Theta\eta^{-1}|\phi_n\rangle\!\rangle$ are linearly
independent \footnote{Though $|\phi_n\rangle$ and
$\Theta\eta^{-1}|\phi_n\rangle\!\rangle$ are linearly
independent, they are not orthogonal, {\it i.e.} 
$\langle\phi_n|\Theta\eta^{-1}|\phi_n\rangle\!\rangle\neq 0$,
in general.}. 
As a result, we have two-fold degeneracy 
in eigenstates of $H$, which is the generalized
Kramers degeneracy. 
 
In general, 
the generalized Kramers partner
$\Theta\eta^{-1}|\phi_n\rangle\!\rangle$ is not coincident with 
$\Theta|\phi_n\rangle$. 
Actually, unlike the TR symmetry with $\Theta^2=-1$, TR symmetry with
$\Theta^2=+1$ does not imply that 
$|\phi_n\rangle$ and $\Theta|\phi_n\rangle$ are linearly independent.
Nevertheless, we can say that if eigenvalues of $H$ are real, the
generalized Kramers partner is essentially the same as $\Theta|\phi_n\rangle$.

To see this, let us consider the eigenstate $|\phi_n\rangle$ satisfying
(\ref{eq:defphin}). Then the pseudo-hermiticity leads to
\begin{eqnarray}
H^{\dagger}\eta|\phi_n\rangle=E_n \eta|\phi_n\rangle. 
\end{eqnarray}
Therefore, $\eta|\phi_n\rangle$ can be expanded as
\begin{eqnarray}
\eta|\phi_n\rangle=\sum_{m}|\phi_m\rangle\!\rangle c_{mn}, 
\label{eq:nexpand}
\end{eqnarray} 
where the sum is taken for $|\phi_m\rangle\!\rangle$'s satisfying
(\ref{eq:defphin2}) and (\ref{eq:bio})  with $E_m^*=E_n$. 
(Note that if there is a degeneracy in the spectrum, we may have multiple such
$|\phi_m\rangle\!\rangle$'s.)
Applying $\langle \phi_m|$ from the left, we obtain
\begin{eqnarray}
c_{mn}=\langle \phi_m|\eta|\phi_n\rangle. 
\end{eqnarray}
Because of the hermiticity of $\eta$, $c_{mn}$ is hermitian for the
indices $m$ and $n$. Thus it can be diagonalized by a unitary matrix $G$
\begin{eqnarray}
\sum_{lk}G_{ml}^{\dagger}c_{lk}G_{kn}=\lambda_m \delta_{mn}, 
\end{eqnarray}
with real $\lambda_m$.
The eigenvalue $\lambda_n$ is not zero because $c_{mn}$ is invertible.
Thus taking the following new bi-orthonormal basis 
\begin{eqnarray}
|\phi_n'\rangle=\sum_m|\phi_m\rangle G_{mn}/\sqrt{|\lambda_n|},
\quad 
|\phi_n'\rangle\!\rangle=\sum_m|\phi_m\rangle\!\rangle 
G_{mn}\sqrt{|\lambda_n|},
\quad 
\langle \phi_n'|\phi_m'\rangle\!\rangle=\delta_{mn},
\label{eq:newbasis}
\end{eqnarray}
we have
\begin{eqnarray}
\eta|\phi_n'\rangle={\rm sgn}(\lambda_n)|\phi_n'\rangle\!\rangle. 
\label{eq:newbaserelation}
\end{eqnarray}
Now suppose that $E_n$ is real, then $E_m$ for $|\phi_m\rangle\!\rangle$
in (\ref{eq:nexpand}) is also real and coincides with $E_n$.
This yields that all $|\phi_m\rangle$'s in the right hand side of the first
equation in (\ref{eq:newbasis}) have the same energy $E_n$.
In other words, $|\phi_n'\rangle$ remains to
be an eigenstate of $H$ with the eigenvalue of $E_n$. 
Applying $\Theta\eta^{-1}$ from the left to both sides of
(\ref{eq:newbaserelation}),  we find that the (generalized) Kramers partner
$\Theta\eta^{-1}|\phi_n'\rangle\!\rangle$ is the same 
as $\Theta|\phi_n'\rangle$ up to an irrelevant overall sign,
\begin{eqnarray}
\Theta|\phi_n'\rangle={\rm sgn}(\lambda_n)
 \Theta\eta^{-1}|\phi'_n\rangle\!\rangle.
\end{eqnarray}
\subsection{Particle/hole symmetry and split-quaternion}
\label{sec:particlehole}

In addition to the TR invariance, we can have the particle/hole symmetry ${\cal C}$
which is antiunitary.
Here we briefly see the split-quaternion structure of particle/hole
symmetric system.

 We say that a Hamiltonian $H$ has the particle/hole symmetry ${\cal C}$ 
if it satisfies
\begin{eqnarray}
{\cal C}H{\cal C}^{-1}=-H. 
\label{eq:particle/hole}
\end{eqnarray}
If we write ${\cal C}$ as ${\cal C}=\Gamma K$ with a unitary operator
$\Gamma$, (\ref{eq:particle/hole}) is recast into
\begin{eqnarray}
\Gamma H \Gamma^{\dagger}=-H^*. 
\end{eqnarray}
One can show that
${\cal C}^2=\pm 1$. 

In a manner similar to the TR symmetry, we have the
split-quaternion structure if ${\cal C}^2=+1$.
The correspondence between the particle/hole symmetry and the
split-quaternion is 
\begin{eqnarray}
(i,{\cal C},i{\cal C}) \leftrightarrow (i,j,k). 
\end{eqnarray}

When $H$ is pseudo-hermitian,
$
H^{\dagger}=\eta H\eta^{-1} 
$
with $\{{\cal C}, \eta\}=0$, we find that eigenstates of $H$ are paired
with eigenvalues $(E_n, -E_n)$.  
Consider 
\begin{eqnarray}
&&H|\phi_n\rangle=E_n|\phi_n\rangle, 
\nonumber
\\
&&H^{\dagger}|\phi_n\rangle\!\rangle
=E_n^{*}|\phi_n\rangle\!\rangle, 
\end{eqnarray}
with $\langle\!\langle \phi_n|\phi_m\rangle=\langle\phi_m|\phi_n\rangle\!\rangle=\delta_{mn}.$
It is found that $|\phi_n\rangle$ and ${\cal
C}\eta^{-1}|\phi_n\rangle\!\rangle$ have the eigenenergies $E_n$ and
$-E_n$, respectively. 
Then, if $\eta$ satisfies $\{\eta, {\cal C}\}=0$, 
we can show
that $|\phi_n\rangle$ and ${\cal
C}\eta^{-1}|\phi_n\rangle\!\rangle$ are linearly independent for any $E_n$,
in a manner similar to Sec.\ref{generalizedKramers}.
Thus, the eigenstates of $H$ are paired.
In particular, if we have a zero energy state with $E=0$, then it should be
degenerated. 

Note that the particle/hole symmetry itself implies that if
$|\phi_n\rangle$ is an eigenstate of $H$ with eigenenergy $E_n$, then
${\cal C}|\phi_n\rangle$ is the one with $-E_n^*$.
For a non-hermitian Hamiltonian, however, this does not always mean
additional pair of states.
Indeed, if $E_n$ is pure imaginary, then $E_n$ is the same as $-E_n^*$,
and $|\phi_n\rangle$ and ${\cal C}|\phi_n\rangle$ can be the
same.
We can also say that if $E_n$ is real, ${\cal C}|\phi_n\rangle$ is
essentially the same as ${\cal C}\eta^{-1}|\phi_n\rangle\!\rangle$:
When the eigenenergies $E_n$ are real, we can take the basis
(\ref{eq:newbasis}) as the eigenstates of $H$, 
which leads to
\begin{eqnarray}
{\cal C}|\phi_n'\rangle={\rm sgn}(\lambda_n){\cal
C}\eta^{-1}|\phi_n'\rangle\!\rangle.
\label{eq:p/hC}
\end{eqnarray}
Thus ${\cal C}|\phi_n'\rangle$ and ${\cal
C}\eta^{-1}|\phi_n'\rangle\!\rangle$ coincide with each other up to a
sign factor.

Formally we can treat the particle/hole symmetry as the
TR symmetry by redefining $H\rightarrow iH$.
In this case, however, the pseudo-hermiticity is replaced by
``pseudo-anti-hermiticity'',
\begin{eqnarray}
H^{\dagger}=-\eta H\eta^{-1}. 
\label{eq:PAH}
\end{eqnarray} 

\subsection{Example: $2\times 2$ matrix}\label{Subsecsplitquat2times2}

In this subsection, we will see the split-quaternion structure in a
concrete example. 
Consider a $2\times 2$ matrix, as the simplest nontrivial Hamiltonian. 
In general, by using the $2\times 2$ unit matrix $1_2$ and the Pauli matrices
$\sigma_i$ $(i=x,y,z)$, 
any $2\times 2$ matrix can be written as  
\begin{eqnarray}
H=h 1_2+\sum_{i=x,y,z}h^i\sigma_i,
\end{eqnarray}
with complex numbers $h$ and $h^{i}$ $(i=x,y,z)$.
Then suppose that $H$ is invariant under the TR symmetry $\Theta=UK$ with
$\Theta^2=+1$.
\begin{eqnarray}
\left[H,\Theta \right]=0.
\end{eqnarray}
$\Theta^2=+1$ implies that $U$ is
a symmetric (unitary) matrix, $U=U^{T}$. 
Following Ref.\cite{Dyson1962}, $U$ can be $U=1_2$ in a proper
basis of the Hamiltonian.
The TR invariance yields 
\begin{eqnarray}
{h}^*={h},
\quad
{h^x}^*={h^x},
\quad
{h^y}^*=-{h^y},
\quad
{h^z}^*={h^z},
\end{eqnarray}
thus, we obtain
\begin{eqnarray}
H=w1_2+x\sigma_x+yi\sigma_y+z\sigma_z,
\label{eq:h2x2}
\end{eqnarray}
with real numbers, $w, x, y, z$.
The split-quaternion structure of $H$ is evident if we notice the following
identification between the Pauli matrices and the basis for the
split-quaternion,
\begin{eqnarray}
(i\sigma_y, \sigma_x, \sigma_z) \leftrightarrow (i,j,k),
\label{eq:paulisplit}
\end{eqnarray}
which reproduces the algebra (\ref{eq:ijk}). 
Thus the TR invariant Hamiltonian (\ref{eq:h2x2}) is a split-quaternion. 

Let us now impose the pseudo-hermiticity.
To satisfy $\{\Theta, \eta\}=0$, the hermitian matrix $\eta$ should be
pure imaginary, $\eta^*=-\eta$.
Thus it can be written as $\eta=c\sigma_y$ with a real number $c$. 
If $H$ is pseudo-hermitian (\ref{pseudohermite}), we obtain 
$x=y=z=0$.
Therefore, the Hamiltonian becomes $H=w 1_2$.
While this Hamiltonian is rather trivial, we have a two-fold
degeneracy. This can be considered as the generalized Kramers
degeneracy explained in Sec.\ref{generalizedKramers}.
In this case, any column vectors
 $|\phi\rangle =(a,b)^{T}$ are
eigenstates of $H$.
The corresponding $|\phi\rangle\!\rangle$ satisfying 
$\langle \phi|\phi \rangle\!\rangle=1$ is
\begin{eqnarray}
|\phi\rangle\!\rangle=
\frac{1}{|a|^2+|b|^2} 
\left(
\begin{array}{c}
a \\
b
\end{array}
\right).
\end{eqnarray}
Thus the generalized Kramers partner,
$\Theta\eta^{-1}|\phi\rangle\!\rangle$, is given by
\begin{eqnarray}
\Theta\eta^{-1}|\phi\rangle\!\rangle=
-\frac{c^{-1}}{|a|^2+|b|^2} 
\sigma_y
\left(
\begin{array}{c}
a^* \\
b^*
\end{array}
\right)
=
\frac{c^{-1}}{|a|^2+|b|^2} 
\left(
\begin{array}{c}
ib^* \\
-ia^*
\end{array}
\right).
\end{eqnarray}
One can easily check that $|\phi\rangle$ and
$\Theta\eta^{-1}|\phi\rangle\!\rangle$ are linearly
independent if $|a|^2+|b|^2\neq 0$. 

Next, impose the pseudo-anti-hermiticity
which implies
$w=0$ in (\ref{eq:h2x2}), so 
\begin{eqnarray}
H=x\sigma_x+yi\sigma_y+z\sigma_z. 
\label{eq:example2}
\end{eqnarray}
The eigenvalues are $E_{\pm}=\pm \sqrt{x^2+z^2-y^2}$.
The corresponding eigenstates $|\phi_{\pm}\rangle$ are given by
\begin{eqnarray}
|\phi_{\pm}\rangle=c_{\pm}
\left(
\begin{array}{c}
x+y \\
\pm E-z
\end{array}
\right),
\quad 
E=\sqrt{x^2+z^2-y^2}, 
\end{eqnarray}
where $c_{\pm}$ are constants.
In accordance with the general argument in Sec.\ref{sec:particlehole},
the eigenstate with the eigenvalue $E$ is paired with  
the one with $-E$. 

We can also check that if we suppose the hermiticity of $H$, instead of the
pseudo-hermiticity or pseudo-anti-hermiticity, 
the split-quaternion structure is replaced by the real structure, 
$y$ in (\ref{eq:h2x2}) being zero, 
thus $H$ reduces to a $2\times 2$ real symmetric matrix.

 \subsection{Integer spin systems}
\label{subsec:trforboson}
In this subsection, we will see the split-quaternion structure of
integer spin systems in which $\Theta^2=+1$.
The spin, $S_i$ $(i=x,y,z)$, changes its sign under the TR
transformation
\begin{equation}
{\Theta}S_i{\Theta}^{-1}=-S_i.
\label{timetransspin}
\end{equation}
Write
$
{\Theta}=U K, 
$
and the condition (\ref{timetransspin}) is written as  
\begin{equation}
U S_x U^{-1}=-S_x,~~U S_y U^{-1}=S_y,~~U S_z U^{-1}=-S_z, 
\end{equation}
where we have assumed the standard matrix realization of $S_i$ in which
only $S_y$ is complex and pure imaginary.
Then, $U$ is given by \footnote{
The form of TR operator depends on the physical meaning of the operator.
 For instance, when $S_i$ denote ``isospin'' labeling two different energy levels, the TR operator $\Theta$ is simply given by $\Theta=K$.}   
\begin{equation}
U=e^{i\pi S_y}, 
\end{equation}
and 
\begin{equation}
\Theta^2=e^{2i\pi S_y}=(-1)^{2S}, 
\label{thetasquaredependingS}
\end{equation}
where $S$ represents the magnitude of the spin. 
Thus, for integer  $S$, $\Theta^2=+1$,  while for  half-integer $S$, 
$\Theta^2=-1$.   
For low spins, 
\begin{subequations}
\begin{align}
S=1/2~&:~ {\Theta}=i\sigma_y K,
\label{defofusualtimerevope} \\
S=1~~~&:~
{\Theta}=
\left(\begin{array}{ccc}
 0 & 0 & 1  \\
0 & -1 & 0 \\
1 & 0 & 0
\end{array}\right) K, \label{theta1matrix}\\
S=3/2~&:~
{\Theta}=
\left(\begin{array}{cc}
 0 & i\sigma_y  \\
i\sigma_y & 0 
\end{array}\right) K. 
\label{defofusualtimerevope32}
\end{align}
\end{subequations}

Let us first consider the $S=1$ system. 
The TR operator is given by (\ref{theta1matrix}). 
Since $U$ is real symmetric matrix, it can be diagonalized by 
an orthogonal matrix $O$, 
\begin{eqnarray}
\Theta=O
\left(
\begin{array}{ccc}
1 &0 &0 \\
0 &-1 &0 \\
0& 0 & -1 
\end{array}
\right)O^{T} K.
\end{eqnarray}
Then $\Theta$ is written as
\begin{eqnarray}
\Theta=OVV^T O^TK=(OV)K (OV)^{\dagger}
\end{eqnarray}
with 
$
V={\rm diag}(1,i,i).
$
Therefore, performing the unitary transformation $OV$ on the basis, $\Theta$
is recast into
$
\Theta=K. 
$

Let us now impose the TR invariance $\Theta$ on the Hamiltonian. 
In general, the Hamiltonian $H$ is written as
\begin{eqnarray}
H=\left(
\begin{array}{cc}
h+h^x\sigma_x+ih^y\sigma_y+h^z \sigma_z & {\bm a}^{\rm T} \\
{\bm b} & c
\end{array}
  \right), 
\end{eqnarray}
where $h$ and $h^i$ $(i=x,y,z)$ are complex numbers, ${\bm a}$ and ${\bm
b}$ are two component complex vectors, and $c$ is a complex number.
Take $[H,\Theta]=0$ with $\Theta=K$, then all the elements of $H$ are real.
In particular, from the correspondence (\ref{eq:paulisplit}), this
implies that the left upper part of $H$ is given by a split-quaternion, 
$\bm{h}^r=w+x\sigma_x+i y\sigma_y+z\sigma_z$ with real
coefficients $w,x,y,z$.  

Next, consider the $S=2$ case, in which $S_y$ is given by 
\begin{eqnarray}
S_y=\frac{i}{2}
\left(
\begin{array}{ccccc}
0 &-2 &0 &0 &0 \\
2 & 0 & -\sqrt{6} & 0 & 0 \\
0 & \sqrt{6} & 0 & -\sqrt{6} & 0 \\
0 & 0 & \sqrt{6}& 0 &-2 \\
0 & 0 & 0 & 2 & 0
\end{array}
\right), 
\end{eqnarray}
and the corresponding TR operator $\Theta$ becomes
\begin{eqnarray}
\Theta=
\left(
\begin{array}{ccccc}
0 &0 &0 &0 &1 \\
0 & 0 &0 & -1 & 0 \\
0 & 0 & 1 & 0 & 0 \\
0 & -1 & 0 & 0 & 0 \\
1 & 0 & 0 & 0 & 0
\end{array}
\right) K. 
\end{eqnarray}
In a manner similar to the above, we have $\Theta=K$
by choosing the basis properly.
The TR invariant Hamiltonian is given by 
\begin{eqnarray}
H=\left(
\begin{array}{ccc}
{\bm h}^{R}_{11} & {\bm h}^{R}_{12}& {\bm a}_1^{\rm T}\\
{\bm h}^{R}_{21} & {\bm h}^{R}_{22} & {\bm a}_2^{\rm T} \\
{\bm b}_1 & {\bm b}_2 & c
\end{array}
  \right), 
\end{eqnarray}
where ${\bm h}^{R}_{IJ}$ $(I,J=1,2)$ are split-quaternions of the form 
\begin{eqnarray}
{\bm h}^{R}_{IJ}=w_{IJ}+x_{IJ}\sigma_x+iy_{IJ}\sigma_y
+z_{IJ} \sigma_z, 
\label{matrixrealsplitquaternion}
\end{eqnarray}
with real coefficients $w_{IJ},x_{IJ},y_{IJ},z_{IJ}$.  
Here ${\bm a}_I$ and ${\bm b}_I$ ($I=1,2$) are real two row vectors, and
$c$ is a real constant.  

In general, for $S=M$ with integer $M$, we can always choose the basis
in which $\Theta$ is given by $\Theta=K$.
Then a TR invariant
Hamiltonian $H$ is given by
\begin{eqnarray}
H=\left(
\begin{array}{cc}
{\bm h}_{IJ}^{R} &{\bm a}_I^{\rm T} \\
{\bm b}_I & c 
\end{array}
\right), 
\label{eq:standard_form}
\end{eqnarray}
where 
${\bm h}_{IJ}^{R}$ ($I, J =1, \cdots, M$) 
are split-quaternions, 
and ${\bm a}_I$ and ${\bm b}_I$ ($I=1,\cdots, M$) are real two row
vectors, and $c$ is a real constant.  

Note that if we impose the hermiticity on the Hamiltonian
(\ref{eq:standard_form}), $H$ reduces to a real symmetric Hamiltonian. 
Thus, our result here is consistent with the known result 
that the hermitian TR invariant systems with integer spins belong to the
orthogonal ensembles \cite{Dyson1962,MehtaBook}. 

\section{Random matrix classification}\label{sectoinrandom}

The idea of random matrix ensembles, pioneered by Wigner and Dyson, is
based on classifying classes of matrices by discrete symmetries
(e.g. see \cite{MehtaBook}). 
Altland and Zirnbauer established the hermitian random matrix theory   
in the context of
superconductivity \cite{Zirnbauer1996,AtlandZirnbauer1997}, which
contains 10 classes. The random matrix classification was also applied
to topological insulators \cite{Schnyderetal2008}. 

Bernard and LeClair extended the random matrix classification for
non-hermitian matrices \cite{Bernard2001}.
In this section, we apply the arguments in
Sec.\ref{splitquaernionsandTreversal} to the framework of the random matrix classification. 

\subsection{Non-hermitian random matrix classification}

Following Ref. \cite{Bernard2001}, 
let us consider discrete symmetries on
non-hermitian random matrices.
Suppose that the discrete symmetries are
implemented by unitary transformations, and the system comes back 
to the original up to a phase factor if they are applied twice.   
Then there are four possible transformations on a non-hermitian 
random matrix $H$
\footnote{The terminology ``symmetry'' is usually used for the operations 
that commute with Hamiltonian, but in this section, ``symmetry'' refers to the
operations (\ref{randommatrixcategory4})-(\ref{randommatrixcategory2}).}:
\begin{subequations}
\begin{align}
& K \quad {\rm sym.}: \quad
H=k H^* k^{-1}, \quad k k^*=\pm { 1}, 
\label{randommatrixcategory4}\\
& Q \quad {\rm sym.}: \quad
H=\epsilon_q 
q H^{\dag} q^{-1}, \quad q^{\dag} q^{-1}={ 1},
\label{randommatrixcategory3}\\
&C \quad {\rm sym.}: \quad
H=\epsilon_c c H^T c^{-1},\quad c^{T} c^{-1}=\pm { 1},
\label{randommatrixcategory1}\\
& P \quad {\rm sym.}: \quad
H=-p H p^{-1},\quad p^2= {1},
\label{randommatrixcategory2}
\end{align} 
\end{subequations}
where $\epsilon_c$ and $\epsilon_q$ are signs, ${\it i.e.}$
 $\epsilon_{c}=\pm 1$, and $\epsilon_{q}=\pm1$, and 
$k$, $q$, $c$, and $p$ are unitary matrices, 
\begin{equation}
kk^{\dag}={ 1},
\quad
qq^{\dag}={1},
\quad 
cc^{\dag}={ 1},
\quad pp^{\dag}={ 1}.
\end{equation}
Demanding that the transformations
(\ref{randommatrixcategory4})-(\ref{randommatrixcategory2})
commute, we have
\begin{eqnarray}
q^{*}=\pm k^{-1}q k^{\dagger -1},
\quad
k^{T}c^{-1}k c^*=\pm 1,
\quad
p^*=\pm k^{-1}p k,
\nonumber\\
q^{T}=\pm c^{\dagger}q^{-1}c,
\quad  
q=\pm p q p^{\dagger},
\quad
c=\pm p c p^{t}.
\label{eq:cd}
\end{eqnarray}

We refer to K symmetry as the TR symmetry.
We can add the minus sign to the right hand side of the first equation
in (\ref{randommatrixcategory4}) by redefining $H\to iH$. 
Thus we can also consider K symmetry as the particle/hole symmetry.

Q symmetry corresponds to the pseudo-hermiticity $(\epsilon_q=1)$ or 
the pseudo-anti-hermiticity $(\epsilon_q=-1)$, defined in the
previous section,  by identifying $q$ with $\eta^{-1}$.
We note that the correspondence is not one-to-one.
While $q$ is a unitary operator, $\eta$ is not always. 
(Both $\eta$ and $q$ are hermitian.)
Thus Q symmetry is a part of the pseudo-(anti-)hermiticity. 

In the case of hermitian matrices, K symmetry is nothing but C
symmetry.
Thus, one often refers to C symmetry as the TR
symmetry if $\epsilon_c=1$ or particle/hole symmetry if $\epsilon_c=-1$.
For non-hermitian matrices, however, they are different. 
Thus, in this paper, we do not refer to C symmetry as the
TR symmetry or particle/hole symmetry. 
C symmetry is obtained by combining
K  and  Q  symmetries. Specifically, from K (\ref{randommatrixcategory4}) and Q (\ref{randommatrixcategory3}) symmetries, C symmetry (\ref{randommatrixcategory1}) is obtained with $c=kq^*$
(up to a phase factor) and $\epsilon_c=\epsilon_q$.

Finally, $P$ symmetry is called the chiral symmetry in literatures.

\subsection{Random matrix classification and split-quaternion}\label{subsect:randomclasssplit}

Write
\begin{eqnarray}
k=U,
\end{eqnarray}
then (\ref{randommatrixcategory4}) gives the TR symmetry with $\Theta=UK$ as 
\begin{eqnarray}
[\Theta, H]=0, 
\quad
\Theta^2=\pm 1. 
\end{eqnarray}
Next write 
\begin{eqnarray}
q=\eta^{-1},
\end{eqnarray}  
then (\ref{randommatrixcategory3}) reads
\begin{eqnarray}
H^{\dagger}=\epsilon_q \eta H \eta^{-1},
\quad
\eta^{\dagger}=\eta.
\end{eqnarray}
Thus Q symmetry with $\epsilon_q=1$ as the
pseudo-hermiticity, and with $\epsilon_q=-1$ as the
pseudo-anti-hermiticity.

If the first equation of (\ref{eq:cd}) holds, a system has both K and Q
symmetries. In terms of $\Theta$ and $\eta$, the commutativity between
K and Q, {\it i.e.} $q^*=\pm k^{-1}qk^{\dagger -1}$, is written as
\begin{eqnarray}
\Theta \eta \mp \eta \Theta=0. 
\end{eqnarray}
Thus K symmetry and Q symmetry are equipped with all the properties
used in the arguments in the previous section.

The arguments in the previous section lead to the following. 
\begin{enumerate}
 \item  \label{firsttheorem} When a non-hermitian matrix has K symmetry with $kk^*=1$, the
       matrix supports the split-quaternion structure.
\item  \label{secondtheorem} 
       If the non-hermitian matrix also has Q symmetry with
       $\epsilon_q=1$ and $q^*=-k^{-1}qk^{\dagger-1}$, at the same time,
       each eigenvalue of the non-hermitian matrix has two-fold
       degeneracy. 
\item \label{thirdtheorem}  
      If the sign of Q symmetry is minus, {\it i.e.} $\epsilon_q=-1$
      (and if $q$ and $k$ satisfy $q^*=-k^{-1}qk^{\dagger-1}$),
      each eigenstate with the eigenvalue $E$ of the non-hermitian matrix 
      has a partner state with $-E$. 
\end{enumerate}
The second result is nothing but the generalized Kramers degeneracy in
Sec.\ref{generalizedKramers}, and the last one comes from 
particle/hole symmetry arguments in Sec.\ref{sec:particlehole}. 

\section{Random matrix class of the $SU(1,1)$ model}\label{randommatrixcategorysu11}

As a concrete realization of the statements in Sec.\ref{subsect:randomclasssplit}, we introduce the $SU(1,1)$ models. 
First consider K symmetry realized by $k=\sigma_x$ with $kk^*=1$.  
Although $k$ can be $k=1_2$ by taking a proper basis\cite{Dyson1962}
as mentioned in Sec. \ref{Subsecsplitquat2times2} and explicitly shown below, 
the present form of $k=\sigma_x$ is convenient to see the $SU(1,1)$ structure.

Any arbitrary $2\times 2$ matrix can be expanded by 2$\times$2 unit and
the Pauli matrices 
\begin{equation}
H=h 1_2+\sum_{i=x,y,z}h^i\sigma_i,  
\end{equation}
where $h$ and $h^{i}$ $(i=x,y,z)$ stand for complex parameters. 
The imposition of K symmetry specifies $h$ and $h^{i}$ as 
\begin{equation}
{h}^*={h},~~~{h^x}^*={h^x},~~~{h^y}^*={h^y},~~~~{h^z}^*=-{h^z}.
\end{equation}
Thus, $h$, $h^x$, $h^y$ are real parameters, while $h^z$ 
a pure imaginary parameter. 
Thus, the Hamiltonian is rewritten as 
\begin{equation}
H=w 1_2+x\sigma_x+y\sigma_y+ i z\sigma_z,  
\end{equation}
with real parameters $w,x,y,z$. 
We further impose Q symmetry with $q=\sigma_z$. 
There are two types of Q symmetry, corresponding to $\epsilon_q=\pm 1$.  

\subsection{$\epsilon_q=+1$ : pseudo-hermiticity}
In this case, the Hamiltonian takes the form of  
\begin{equation}
H=w 1_2. 
\label{2times2unitmat}
\end{equation}
The Kramers degeneracy of this Hamiltonian was discussed in
Sec.\ref{Subsecsplitquat2times2}.
Indeed, K symmetry here corresponds to $\Theta=\sigma_x K$.
Thus by applying the following unitary transformation, 
\begin{eqnarray}
\Theta\rightarrow V^{\dagger}\Theta V,
\quad
V=
\frac{1}{\sqrt{2}}
\left(
\begin{array}{cc}
1 & -i\\
1 &  i
\end{array}
\right), 
\end{eqnarray}
$\Theta$ reduces to the one in Sec.\ref{Subsecsplitquat2times2},
$\Theta=K$.
At the same time, in this unitary transformation, $q$ and $H$ become
\begin{eqnarray}
q\rightarrow V^{\dagger}\sigma_z V=\sigma_y, 
\quad\quad
H\rightarrow V^{\dagger}w 1_2V=w1_2,
\end{eqnarray}
which are also the same as those in
Sec.\ref{Subsecsplitquat2times2}.
Thus the degeneracy here can be understood as a consequence of the
generalized Kramers. 
\subsection{$\epsilon_q=-1$ : pseudo-anti-hermiticity}
When $\epsilon_q=-1$,
the Hamiltonian becomes 
\begin{equation}
H=x\sigma_x+y\sigma_y+ iz\sigma_z. 
\end{equation}
From the general theorem 3 in Sec.\ref{subsect:randomclasssplit},  $H$ is
expected to have a partner state with $E$ and $-E$.
Actually, the Hamiltonian has eigenvalues $\pm\sqrt{x^2+y^2-z^2}$, thus
the energy eigenvalues are paired.
The eigenvalues are real when $x^2+y^2-z^2\ge 0$, which corresponds to
the model in quantum optics \cite{BenAryeh2004}.
In this case, the constant energy surface in the parameter space is a one-leaf
hyperboloid, $H^{1,1}$.
On the other hand, when $x^2+y^2-z^2\le 0$, the constant energy surface
in the parameter space is described by the two-leaf hyperboloid,
$H^{2,0}$.
Here we will consider the properties of the latter.  

The corresponding Hamiltonian takes the pure imaginary eigenvalues with
opposite sign. Therefore, we transform $H$ into $iH$, then deal with the
following Hamiltonian,
\begin{equation}
H=-x\tau_x-y\tau_y+z\tau_z= 
\left(\begin{array}{cc}
z & -ix-y \\
-ix+y & -z
\end{array}\right). 
\label{SU11H}
\end{equation}
Here $\tau_i$ $(i=x,y,z)$ are ``Pauli matrices'' of $SU(1,1)$: 
\begin{equation}
\tau_x=i\sigma_x,~~\tau_y=i\sigma_y,~~\tau_z=\sigma_z.  
\end{equation}
 (See Appendix \ref{appensu11spinquadrupole}, also.) 
As a consequence of the transformation $H\rightarrow iH$, the TR
symmetry in the original Hamiltonian is converted into 
the particle/hole symmetry 
 \begin{equation}
 \mathcal{C}H\mathcal{C}^{-1}=-H,
\quad
\mathcal{C} = \sigma_x \cdot K,
\label{psedotimereversal12}
 \end{equation}
and the pseudo-anti-hermiticity becomes the pseudo-hermiticity,
\begin{eqnarray}
H^{\dagger}=\eta H \eta^{-1}, \quad \eta=\sigma_z. 
\end{eqnarray} 

Let us consider situations where all the eigenvalues of 
the Hamiltonian (\ref{SU11H}) are real.
Then, the parameters $x,y,$ and $z$ give coordinates on 
a two-leaf hyperboloid $H^{2,0}$:
\begin{equation}
 {x}^2+{y}^2-{z}^2=-r^2 \le 0,
\label{conditionH20}
\end{equation}
where $r$ is a real positive constant. Since $x$ and $y$ are real, $z$ is taken either $z\ge r$ (upper leaf) or $z\le -r$ (lower leaf).   
The eigenvalues of the $SU(1,1)$  Hamiltonian (\ref{SU11H}) are given by 
\begin{equation}
E_{\pm}=\pm r.
\label{energyeigensu11}
\end{equation}
On the upper leaf ($z\ge r$), the eigenvectors are given by  
\begin{equation}
|\phi_{+}\rangle=\frac{1}{\sqrt{2r(r+ z)}}
\left(\begin{array}{c}
r+ z \\
y-ix
\end{array}\right),~~
|\phi_{-}\rangle=\frac{1}{\sqrt{2r(r+z)}}
\left(\begin{array}{c}
y+ix \\
r+z
\end{array}\right),  
\label{biortho1phi}
\end{equation}
which satisfy 
\begin{equation}
\langle\phi_{\pm}|\eta|\phi_{\pm}\rangle=
\langle\phi_{\pm}|\sigma_z |\phi_{\pm}\rangle=\pm 1. 
\label{normalizationforphipm}
\end{equation}

Meanwhile, the eigenvectors of the hermite-conjugate Hamiltonian 
\begin{equation}
H^{\dagger}=x\tau_x+y\tau_y+z\tau_z=
\left(\begin{array}{cc}
z & ix+y\\
ix-y & -z
\end{array}\right),
\label{HermiteSU11hamiltonian}
\end{equation}
are 
\begin{equation}
|\varphi_{+}\rangle\!\rangle=\frac{1}{\sqrt{2r(r+z)}}
\left(\begin{array}{c}
r+z \\
-y+ix
\end{array}\right),~~
|\varphi_{-}\rangle\!\rangle=\frac{1}{\sqrt{2r(r+ z)}}
\left(\begin{array}{c}
-y-ix \\
r+z
\end{array}\right), 
\label{biortho2varphi}
\end{equation}
which satisfy  
\begin{equation}
\langle\!\langle\varphi_{\pm}|\eta^{-1} |\varphi_{\pm}\rangle\!\rangle=
\langle\!\langle\varphi_{\pm}|\sigma_z |\varphi_{\pm}\rangle\!\rangle=\pm 1. 
\end{equation}
The bi-orthonormal bases, $|\phi_{\pm}\rangle$ and $|\varphi_{\pm}\rangle\!\rangle$, satisfy 
\begin{equation}
\langle \phi_{m}| \varphi_n\rangle\!\rangle =\delta_{mn}, ~~~~~  
\sum_{m=\pm}|\phi_{m}\rangle\langle\!\langle\varphi_m|
=|\phi_+\rangle\langle\!\langle\varphi_+|
+|\phi_-\rangle\langle\!\langle\varphi_-|=1_2,   
\end{equation}
and  are related as \footnote{Similarly, for the lower-leaf ($z\le -r$), 
the eigenvectors are   
\begin{equation}
|\phi'_{+}\rangle=\frac{1}{\sqrt{2r(r-z)}}
\left(\begin{array}{c}
-y-ix \\
r-z
\end{array}\right),~~
|\phi'_{-}\rangle=\frac{1}{\sqrt{2r(r-z)}}
\left(\begin{array}{c}
r-z\\
-y+ix
\end{array}\right), 
\label{biortho1phiupper}
\end{equation}
and 
\begin{equation}
|\varphi'_{+}\rangle\!\rangle=\frac{1}{\sqrt{2r(r-z)}}
\left(\begin{array}{c}
y+ix \\
r-z
\end{array}\right),~~
|\varphi'_{-}\rangle\!\rangle=\frac{1}{\sqrt{2r(r-z)}}
\left(\begin{array}{c}
r-z \\
y-ix
\end{array}\right). 
\label{biortho2varphiupper}
\end{equation}
They satisfy 
\begin{equation}
\langle{\phi_{\pm}'}|\sigma_z|\phi_{\pm}'\rangle
=\langle\!\langle{\varphi_{\pm}'}|\sigma_z|\varphi_{\pm}'\rangle\!\rangle
=\mp 1,
~~
\langle \phi_m'|\varphi'_n\rangle\!\rangle=\delta_{mn}. 
\label{normalizationforupper}
\end{equation}
} 
 \begin{equation}
|\phi_{\pm}\rangle~\overset{\eta}\longrightarrow~\eta|\phi_{\pm}\rangle
=\sigma_z|\phi_{\pm}\rangle=\pm |\varphi_{\pm}\rangle\!\rangle.
\label{etatransforofphi}
 \end{equation}

From the general arguments in Sec. \ref{sec:particlehole}, the particle/hole pair of $|\phi_{m}\rangle$ is given by $\mathcal{C}\eta^{-1}\;|\varphi_m\rangle\!\rangle$. 
 With $\eta=\sigma_z$ and $\mathcal{C}=\sigma_x K$, the particle/hole pair of $|\phi_{\pm}\rangle$ is given by 
\begin{equation} 
\mathcal{C}\eta^{-1}\; |\varphi_{\pm}\rangle\!\rangle=
 -i\sigma_y K |\varphi_{\pm}\rangle\!\rangle=\pm |\phi_{\mp}\rangle.
\end{equation}
Thus, the particle/hole pair of $|\phi_+\rangle$ is $|\phi_-\rangle$. 
Indeed, $|\phi_+\rangle$ and $|\phi_-\rangle$ are eigenstates of $H$
with opposite energies, $E_{+}=r$ and $E_-=-r$. 
Though $|\phi_+\rangle$ and $|\phi_-\rangle$ are not orthogonal in the usual
sense, {\it
i.e.} $\langle \phi_+|\phi_-\rangle\neq 0$, they are 
linearly independent. They are 
orthogonal in the sense of pseudo-inner product: 
\begin{equation}
\langle \phi_-|\sigma_z|\phi_+\rangle=0. 
\end{equation}
Applying $\mathcal{C}\eta^{-1}$ from the left to the both sides of
(\ref{etatransforofphi}),
$\eta|\phi_{\pm}\rangle=\pm|\varphi_{\pm}\rangle\!\rangle$, we have
\begin{equation}
\mathcal{C}|\phi_{\pm}\rangle 
= \pm \mathcal{C}\eta^{-1}|\varphi_{\pm}\rangle\!\rangle.
\end{equation}
Therefore, $\mathcal{C}\eta^{-1}|\varphi_{\pm}\rangle\!\rangle$  is equal to $\mathcal{C}|\phi_{\pm}\rangle$ up to sign. 
(See also (\ref{eq:p/hC}) in Sec.\ref{sec:particlehole}.)
Indeed, 
\begin{eqnarray}
|\phi_{\pm}\rangle~\overset{\mathcal{C}}\longrightarrow~|\phi_{\mp}\rangle
=\mathcal{C}|\phi_{\pm}\rangle. 
\end{eqnarray}
Thus, we find that the particle/hole pair of $|\phi_{\pm}\rangle$ is simply given by $\mathcal{C}|\phi_{\pm}\rangle$ in the present case.  

The relations of bi-orthonormal bases,
$|\phi_{\pm}\rangle$ and $|\varphi_{\pm}\rangle\!\rangle$, are
summarized in Fig.\ref{relationssu11eigenvector}. 
\begin{figure}[tbph]\begin{center}
\includegraphics*[width=39mm]{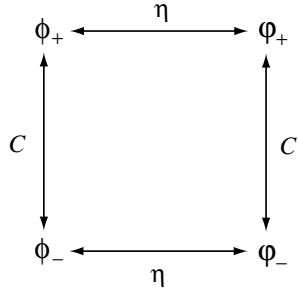}
\end{center}
\caption{The relations between the eigenvectors of the $SU(1,1)$ model.}
\label{relationssu11eigenvector}
\vspace{-3mm}
\end{figure}

\section{Random matrix class of $SO(3,2)$ model}
\label{randommatrixcategoryso32}

As a non-trivial realization of the generalized Kramers,  we introduce  $SO(3,2)$ model. 
First consider K symmetry realized by 
\begin{equation}
k=
\begin{pmatrix}
\sigma_x & 0 \\
0 & \sigma_x 
\end{pmatrix}, 
\label{so32k}
\end{equation}
which satisfies $kk^*=1$. 
Any arbitrary $4\times 4$ matrix can be expanded as 
\begin{equation}
H=h\;{1_4}+\sum_{a=1}^5h^a\gamma_a+\sum_{a<b=1}^5 h^{ab}\gamma_{ab},  
\end{equation}
where $1_4$ is $4\times 4$ unit matrix, $h$, $h^a$ and $h^{ab}$  stand for complex parameters, $\gamma_a$ denote $SO(3,2)$ gamma matrices  
\begin{equation}
\gamma_1=\begin{pmatrix}
0 & \sigma_x \\
-\sigma_x & 0 
\end{pmatrix},~\gamma_2=\begin{pmatrix}
0 & \sigma_y \\
-\sigma_y & 0 
\end{pmatrix},~\gamma_3=\begin{pmatrix}
0 & -i\sigma_z \\
i\sigma_z & 0 
\end{pmatrix},~\gamma_4=\begin{pmatrix}
0 & 1_2\\
1_2 & 0 
\end{pmatrix},~\gamma_5=\begin{pmatrix}
1_2 & 0   \\
0 & -1_2 
\end{pmatrix},\label{so32gammamain}
\end{equation}
and $\gamma_{ab}$ are $SO(3,2)$ generators constructed by  
\begin{equation}
\gamma_{ab}=\frac{1}{4 i}[\gamma_a,\gamma_b]. 
\label{gammamaingene}
\end{equation}
The sixteen matrices, ${1_4}$, $\gamma_a$, $\gamma_{ab}$, amount to complete matrix bases that span arbitrary $4\times 4$ matrix. 
The imposition of K symmetry specifies $h$ and $h^{a}$ as real parameters
and $h^{ab}$ as pure imaginary parameters, 
\begin{equation}
{h}^*={h},~~~{h^a}^*={h^a},~~~{h^{ab}}^*=-{h^{ab}}.
\end{equation}
Thus, the Hamiltonian becomes  
\begin{equation}
H=w\;{1}_4+\sum_{a=1}^5 x^a\gamma_a+i\sum_{a<b=1}^5 x^{ab}\gamma_{ab},  
\end{equation}
with real parameters, $w$, $x^{a}$ and $x^{ab}$. 
We further impose Q symmetry 
\begin{equation}
q=\begin{pmatrix}
\sigma_z & 0 \\
0 & \sigma_z 
\end{pmatrix}, 
\label{so32q}
\end{equation}
which satisfies $q^*=-k^{-1}qk^{\dagger -1}$.
According to two types of Q symmetry, $\epsilon_q=\pm 1$, the
Hamiltonian takes two different forms shown in Sec.\ref{sec:ph} and
Sec.\ref{sec:pah}, respectively. 

\subsection{$\epsilon_q=+1$: pseudo-hermiticity}
\label{sec:ph}

In this case, the Hamiltonian becomes \footnote{This is the first nontrivial form of self-dual real split-quaternion matrix. See Appendix \ref{appendsplit} for details.}
\begin{equation}
H=w\;{1}_4+\sum_{a=1}^5 x^a\gamma_a. 
\label{gammaso33hamil}
\end{equation}
From the general theorem \ref{secondtheorem} in Sec.\ref{subsect:randomclasssplit},  
this Hamiltonian  is expected to exhibit Kramers degeneracy. 
The Hamiltonian is invariant under the time reversal symmetry 
\begin{equation}
\Theta H\Theta^{-1}=H,
\end{equation}
where 
\begin{equation}
\Theta=
\begin{pmatrix}
\sigma_x & 0 \\
0 & \sigma_x
\end{pmatrix} \cdot K, 
\label{so32thetafirst}
\end{equation}
with $\Theta^2=+1$.

The $SO(3,2)$ Hamiltonian (\ref{gammaso33hamil}) is rewritten as        
\begin{align}
H=
\left(\begin{array}{cc}
x^5 1_2 & x^4 1_2 -ix^i\tau_i\\
x^4 1_2 +ix^i\tau_i & -x^5 1_2
\end{array}\right), 
\label{SO32hamiltonian}
\end{align}
where $x^i\tau_i\equiv x\tau_x+y\tau_y+z\tau_z$.  
($\sum_{i=1}^3x^i\tau_i$ will be abbreviated as $x^i\tau_i$ hereafter.) 
The eigenvalues of $H$ are derived as
$E_{\pm}=\pm\sqrt{-(x^1)^2-(x^2)^2+(x^3)^2+(x^4)^2+(x^5)^2}$.
Note that we have two-fold degeneracy in the spectrum, which
comes from the generalized Kramers theorem mentioned in
Sec.\ref{generalizedKramers}.

Let us consider situations where all the eigenvalues 
of (\ref{SO32hamiltonian}) are real:
\begin{equation}
 (x^1)^2+(x^2)^2-(x^3)^2-(x^4)^2-(x^5)^2 =-r^2\le 0, 
\label{condso32parameters} 
\end{equation}
where $r$ is a real positive constant. 
(When $x^4=x^5=0$, the $SO(3,2)$ model is reduced to two
independent  $SU(1,1)$ models.) The eigenvalues 
of the $SO(3,2)$ model (\ref{SO32hamiltonian}) are
\begin{equation}
E_{\pm}=\pm r.   
\end{equation}
Here $E_+$ and $E_-$ are doubly degenerate, respectively.   
For $x^5 > -r$, the eigenvectors are given by 
\begin{align}
&|\psi_{+\alpha}\rangle=\frac{1}{\sqrt{2r(r+x^5)}}
\left(\begin{array}{c}
(r+x^5) \phi_{\alpha} \\
(x^4+ix^i\tau_i) \phi_{\alpha}
\end{array}\right), \nonumber\\
&|\psi_{-\alpha}\rangle=\frac{1}{\sqrt{2r(r+x^5)}}
\left(\begin{array}{c}
(-x^4+ix^i\tau_i) \phi_{\alpha} \\
(r+x^5) \phi_{\alpha} 
\end{array}\right), 
\label{biortho1psi}
\end{align}
where $\phi_{\alpha}$ $(\alpha=\pm)$ represent two-component spinors that account for double degeneracy. Take $\phi_{\pm}$ as 
\begin{equation}
\phi_{+}=
\left(\begin{array}{c}
1 \\
0
\end{array}\right),~~~
\phi_{-}=
\left(\begin{array}{c}
0 \\
1
\end{array}\right). 
\end{equation}

The hermitian conjugate of the Hamiltonian (\ref{SO32hamiltonian}) is  
\begin{align}
H^{\dagger}=\left(\begin{array}{cc}
x^5 1_2 & x^4 1_2 -ix^i\tau_i^{\dagger} \\
 x^4 1_2 +ix^i\tau_i^{\dagger} & -x^5 1_2
\end{array}\right). 
\label{SO32hamiltoniandagger}
\end{align}
Since 
\begin{equation}
{H^{\dagger}}^2=H^2=r^2 1_4,  
\end{equation}
the eigenvalues of $H^{\dagger}$ are also given by $\pm r$, and the corresponding eigenvectors are \footnote{For $x^5 < r$, the eigenvectors are
\begin{align}
&|\psi'_{+\alpha}\rangle=\frac{1}{\sqrt{2r(r-x^5)}}   
\left(
\begin{array}{c}
(-x^4+ix^i\tau_i)\phi_{\alpha} \\
 -(r-x^5)\phi_{\alpha}
 \end{array}\right), 
\nonumber\\
&|\psi'_{-\alpha}\rangle=\frac{1}{\sqrt{2r(r-x^5)}}   
\left(
\begin{array}{c}
 (r-x^5)\phi_{\alpha}\\
-(x^4+ix^i\tau_i)\phi_{\alpha}
 \end{array}\right).  
\end{align}
$|\psi_{\pm\alpha}\rangle$ and $|\psi'_{\pm\alpha}\rangle$ are related
by the $SU(1,1)$ transformation 
\begin{equation}
g_{\pm}=\frac{1}{\sqrt{r^2-(x^5)^2}} (-x^4\pm ix^i\tau_i), 
\label{su11groupelements}
\end{equation}
where $g_-={g_+}^{-1}$. 
The eigenvectors of $H^{\dagger}$ are 
\begin{align}
&|\chi'_{+\alpha}\rangle\!\rangle=\frac{1}{\sqrt{2r(r-x^5)}}   
\left(
\begin{array}{c}
(-x^4+ix^i\tau_i^{\dagger})\phi_{\alpha} \\
 -(r-x^5)\phi_{\alpha}
 \end{array}\right), 
\nonumber\\
&|\chi'_{-\alpha}\rangle\!\rangle=\frac{1}{\sqrt{2r(r-x^5)}}   
\left(
\begin{array}{c}
(r-x^5)\phi_{\alpha}\\
-(x^4+ix^i\tau_i^{\dagger})\phi_{\alpha}
 \end{array}\right).  
\end{align}
$|\chi_{\pm\alpha}\rangle\!\rangle$ and
$|\chi'_{\pm\alpha}\rangle\!\rangle$ are related by the $SU(1,1)$
transformation  
\begin{equation}
g'_{\pm}=\frac{1}{\sqrt{r^2-(x^5)^2}} (-x^4\pm ix^i\tau_i^{\dagger}),  
\end{equation}
where $g'_{-}={g'_+}^{-1}.$} 
\begin{align}
&|\chi_{+\alpha}\rangle\!\rangle=\frac{1}{\sqrt{2r(r+x^5)}}
\left(\begin{array}{c}
(r+x^5) \phi_{\alpha} \\
(x^4+ix^i\tau_i^{\dagger}) \phi_{\alpha}
\end{array}\right), \nonumber\\
&|\chi_{-\alpha}\rangle\!\rangle=\frac{1}{\sqrt{2r(r+x^5)}}
\left(\begin{array}{c}
(-x^4+ix^i\tau_i^{\dagger}) \phi_{\alpha}\\
(r+x^5) \phi_{\alpha} 
\end{array}\right).
\label{biortho2chi}
\end{align}
$|\psi_{m\alpha}\rangle$ and
$|\chi_{m\alpha}\rangle\!\rangle$ are related as
\begin{equation}
|\psi_{m\alpha}\rangle~~\overset{\eta}\longrightarrow ~~
\eta|\psi_{m\alpha}\rangle=\alpha|\chi_{m\alpha}\rangle\!\rangle. 
\label{etatransformpsi}
\end{equation}
With (\ref{biortho1psi}) and (\ref{biortho2chi}), it is straightforward
to confirm
that $|\psi_{m\alpha}\rangle$ and $|\chi_{m\alpha}\rangle\!\rangle$
indeed constitute the bi-orthonormal basis:
\begin{eqnarray}
\langle\psi_{m\alpha}|\chi_{n\beta}\rangle\!\rangle
&=&\delta_{mn}\delta_{\alpha\beta}, 
\\  
\sum_{m,\alpha=+,-}|\psi_{m\alpha}\rangle\langle\!\langle\chi_{m\alpha}|
&=&|\psi_{++}\rangle\langle\!\langle\chi_{++}|+ |\psi_{+-}\rangle
\langle\!\langle\chi_{+-}|  + |\psi_{-+}\rangle\langle\!\langle\chi_{-+}|
+|\psi_{--}\rangle\langle\!\langle \chi_{--}|=1_4. 
\nonumber
\end{eqnarray}
With (\ref{etatransformpsi}), the time-independent inner products induced by
$\eta$ are given as
\begin{equation}
\langle \psi_{m\alpha}|\eta|\psi_{n\beta}\rangle
=\langle\!\langle\chi_{m\alpha}|\eta|\chi_{n\beta}\rangle\!\rangle
=\alpha\delta_{mn}\delta_{\alpha\beta}. 
\end{equation}
Note the sign of these products depends on their ``spin'' directions.  

As mentioned above, 
two-fold degeneracy is a consequence of the generalized Kramers theorem.
The generalized Kramers pair of $|\psi_{m\alpha}\rangle$ 
is given by $\Theta\eta^{-1}|\chi_{m\alpha}\rangle\!\rangle$. Here,
$\eta$ and $\Theta$ are given by $\eta=q$ (\ref{so32q}) and $\Theta$
(\ref{so32thetafirst}).
 Therefore, the generalized Kramers pair is derived as 
\begin{equation} 
\Theta\eta^{-1}|\chi_{m\alpha}\rangle\!\rangle=-
\begin{pmatrix}
i\sigma_y & 0 \\
0 & i\sigma_y
\end{pmatrix}K|\chi_{m\alpha}\rangle\!\rangle
= {\alpha}|\psi_{m\bar{\alpha}}\rangle,   
\end{equation}
where $\bar{\alpha}$ is the opposite spin of $\alpha$; $\overline{\alpha}=-\alpha$, $i.e.$ $\overline{(+)}=-$, $\overline{(-)}=+$.
Thus, $|\psi_{m\alpha}\rangle$ and
$|\psi_{m\bar{\alpha}}\rangle$ 
are indeed the generalized Kramers pair.
 Notice the ``spins'' $\alpha$ of the generalized Kramers pair are opposite 
to each other.  
Though they are not orthogonal in the ordinary sense, they are linearly
independent. 
In the present case, they are orthogonal in the following inner product:
\begin{equation}
\langle \psi_{m\bar{\alpha}}|\eta|\psi_{m\alpha}\rangle=0.
\end{equation}
Since we have the relation (\ref{etatransformpsi}) which corresponds to 
(\ref{eq:newbaserelation}) with real eigenenergies, the generalized Kramers pair  $\Theta\eta^{-1}|\chi_{m\alpha}\rangle\!\rangle$ is equivalent to $\Theta|\psi_{m\alpha}\rangle$ (see Sec.\ref{generalizedKramers}). Indeed, 
\begin{equation}
|\psi_{m\alpha}\rangle~\overset{\Theta}\longrightarrow ~
\Theta|\psi_{m\alpha}\rangle=|\psi_{m\bar{\alpha}}\rangle. 
\end{equation}
Thus, we find the generalized Kramers pair is simply given by  
$\Theta|\psi_{m\alpha}\rangle$.  

The relations between $|\psi_{m\alpha}\rangle$ and
$|\chi_{m\alpha}\rangle\!\rangle$ are summarized in
Fig.\ref{relationsso32eigenvector}. 
\begin{figure}[tbph]\begin{center}
\includegraphics*[width=39mm]{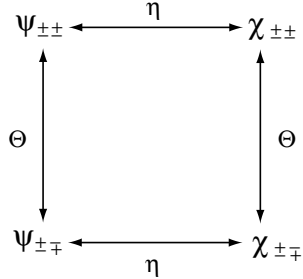}
\end{center}
\caption{The relations between the eigenvectors of the $SO(3,2)$ model.}
\label{relationsso32eigenvector}
\vspace{-3mm}
\end{figure}

\subsection{$\epsilon_q=-1$: pseudo-anti-hermiticity}
\label{sec:pah}
When $\epsilon_q=-1$, 
the Hamiltonian takes the form of 
\begin{equation}
H=i\sum_{a<b=1}^5x^{ab}\gamma_{ab}. 
\end{equation}
The Hamiltonian is invariant under the particle/hole symmetry 
\begin{equation}
\mathcal{C} H\mathcal{C}^{-1}=-H,
\end{equation}
where 
\begin{equation}
\mathcal{C}=
\begin{pmatrix}
\sigma_x & 0 \\
0 & \sigma_x
\end{pmatrix} \cdot K, 
\end{equation}
with $\mathcal{C}^2=+1$.
From the theorem \ref{thirdtheorem} in Sec.\ref{subsect:randomclasssplit},  
we expect $H$ has partner states whose energies are  $E$ and $-E$. Indeed, an explicit calculation shows  the Hamiltonian has two paired states with energies $(E_1,-E_1)$ and $(E_2,-E_2)$ (in general $E_1\neq E_2$)\footnote{The expressions of $E_1$ and $E_2$ are rather lengthy, so we omit their explicit formulae.}. 
\subsection{Realization as $J=3/2$ $SU(1,1)$ quadrupole model}\label{subsec:realizationasj=3/2}

In Ref.\cite{Avron1988}, Avron et al. demonstrated that the $S=3/2$ quadrupole Hamiltonian can be expressed by an $SO(5)$ Hamiltonian. 
Here, we demonstrate how such arguments are generalized to the present non-hermitian case.  
The correspondences between $SU(2)$ and $SO(5)$; and  $SU(1,1)$ and $SO(3,2)$  suggest that the $SU(1,1)$ spin $3/2$ quadrupole  Hamiltonian may be expressed by  an $SO(3,2)$ Hamiltonian.   
The $SU(1,1)$ spin ${J_i}$ $(i=x,y,z)$ are defined so as to satisfy  
\begin{equation}
[J_i,J_j]=i\epsilon_{ijk}J^k, 
\end{equation}
where $\epsilon_{ijk}$ denote the totally antisymmetric 3-rank tensor 
with $\epsilon_{xyz}=1$, and $J^i=(J_x,J_y,-J_z)$.  
With  real $3\times 3$ quadrupole coefficients $Q^{ij}$, we introduce  
the $SU(1,1)$ quadrupole Hamiltonian as 
\begin{equation}
H(Q)=\sum_{i,j=x,y,z} Q^{ij}J_i J_j. 
\label{su11quadrupolehamil1st}
\end{equation}
The $SU(1,1)$ quadrupole Hamiltonian is invariant under the $SU(1,1)$ spin flipping transformation 
\begin{equation}
{J_i}\rightarrow - {J_i}.  
\end{equation}
The five basis elements of $Q^{ij}$ are taken as 
\begin{eqnarray}
&~~~Q_1=\frac{1}{\sqrt{3}}\left(\begin{array}{cccc}
0 & 0 & 1 \\
0 & 0 & 0 \\
1 & 0 & 0
\end{array}\right),~~~~Q_2=\frac{1}{\sqrt{3}}\left(\begin{array}{cccc}
0 & 0 & 0 \\
0 & 0 & 1 \\
0 & 1 & 0
\end{array}\right),~~
&Q_3=\frac{1}{\sqrt{3}}\left(\begin{array}{cccc}
1 & 0 & 0 \\
0 & -1 & 0 \\
0 & 0 & 0
\end{array}\right),\nonumber\\
&Q_4=\frac{1}{\sqrt{3}}\left(\begin{array}{cccc}
0 & 1 & 0 \\
1 & 0 & 0 \\
0 & 0 & 0
\end{array}\right),~~~~Q_5=\frac{1}{3}\left(\begin{array}{cccc}
1 & 0 & 0 \\
0 & 1 & 0 \\
0 & 0 & 2
\end{array}\right), 
\label{qforso32}
\end{eqnarray}
which are orthonormal \footnote{
Note that $Q_5$ is different from the traceless quadrupole matrix,   
 $\frac{1}{3}\begin{pmatrix} 
-1 & 0 & 0 \\
0 & -1 & 0 \\
0 & 0 & 2
\end{pmatrix}$, used in Ref.\cite{Avron1988}.}   
\begin{equation}
(Q_{a},Q_{b})=\frac{3}{2}Tr(Q_{a}Q_{b})=\delta_{ab}. 
\end{equation}
With the use of $Q_a$, an arbitrary quadrupole matrix is expanded as 
\begin{equation}
Q=\sum_{a=1}^5 x^a Q_a,  
\end{equation}
where $x^a$ are real, 
and the $SU(1,1)$ quadrupole Hamiltonian (\ref{su11quadrupolehamil1st}) is expressed as    
\begin{equation}
H(Q)=\sum_{a=1}^5 x^a H(Q_a)
\end{equation}
where  
\begin{equation}
H(Q_a)=\sum_{i,j=x,y,z}{(Q_{a})}^{ij} J_i J_j.
\label{quadrupolehamitlnians}
\end{equation}
More explicitly, they are 
\begin{align}
&H(Q_1)=\frac{1}{\sqrt{3}}\{J_x,J_z\},~~~~~H(Q_2)=\frac{1}{\sqrt{3}}\{J_y,J_z\},
~~~~~H(Q_3)=\frac{1}{\sqrt{3}}(J_x^2-J_y^2),\nonumber\\
&H(Q_4)=\frac{1}{\sqrt{3}}\{J_x,J_y\},~~~~~H(Q_5)=\frac{1}{3}(J_x^2+J_y^2+2J_z^2). 
\label{quadrupolehamcomp}
\end{align}

In particular, for  $J=3/2$, 
the $SU(1,1)$ spin is given by $4\times 4$ matrices  
\begin{align}
J_x=\frac{1}{2}
\left(\begin{array}{cccc}
0 & \sqrt{3}i  & 0 & 0 \\
\sqrt{3}i & 0 & 2i  & 0 \\
0 & 2i & 0 & \sqrt{3}i \\
0 & 0 & \sqrt{3}i & 0 
\end{array}\right), ~
J_y=\frac{1}{2}
\left(\begin{array}{cccc} 
0 & \sqrt{3}  & 0 & 0 \\
-\sqrt{3} & 0 & 2  & 0 \\
0 & -2 & 0 & \sqrt{3} \\
0 & 0 & -\sqrt{3} & 0 
\end{array}\right),~ J_z=\frac{1}{2}
\left(\begin{array}{cccc}
3 & 0 & 0 & 0 \\
0 & 1 & 0 & 0 \\
0 & 0 & -1 & 0 \\
0 & 0 & 0 & -3
\end{array}\right). 
\label{j32su11mat}
\end{align}
The spin flipping operator $\Theta'$  satisfying     
\begin{equation}
 {\Theta'} J_i{\Theta'}^{-1}=-J_i,  
\label{timereversaltransJ}
\end{equation}
 is given by   
\begin{equation}
\Theta'=
\left(\begin{array}{cc}
 0 & \sigma_x  \\
\sigma_x & 0 
\end{array}\right)\cdot K.  
\label{thetadash}
\end{equation}
Since $\Theta'$ satisfies 
\begin{equation}
{\Theta'}^2=+1,  
\end{equation}
$\Theta'$ can be regarded as the TR operator for ${\Theta'}^2=+1$. 
(See Appendix \ref{appensu11spinquadrupole}, for more details about the $
SU(1,1)$ spin flipping operator.) 
Since $H(Q_a)$  are the quadratic forms  of the $SU(1,1)$ spins (\ref{quadrupolehamcomp}), they are invariant under the $SU(1,1)$ spin flipping operation. 
By substituting (\ref{j32su11mat}) into (\ref{quadrupolehamcomp}),
$H(Q_a)$ are explicitly derived as 
\begin{equation}
H(Q_{a})=\gamma'_{a},
\end{equation}
with $\gamma_a'$ being $SO(3,2)$ gamma matrices  
\begin{align}
&\gamma'_1=
\left(\begin{array}{cc}
i\sigma_x & 0 \\
0 & -i\sigma_x
\end{array}\right),~~~~ \gamma'_2=
\left(\begin{array}{cc}
i\sigma_y & 0 \\
0 & -i\sigma_y
\end{array}\right), ~~
\gamma'_3=
\left(\begin{array}{cc}
0 & -1_2 \\
-1_2 & 0 
\end{array}\right),\nonumber\\
&\gamma'_4=
\left(\begin{array}{cc}
0 & i1_2 \\
-i1_2 & 0 
\end{array}\right),~~~~~
\gamma'_5=
\left(\begin{array}{cc}
\sigma_z & 0 \\
0 & -\sigma_z
\end{array}\right).
\label{gammadash}
\end{align}
Therefore, the $J=3/2$ $SU(1,1)$ quadrupole model is expressed by 
\begin{equation}
H=\sum_{a=1}^5 x^a {\gamma'}_a, 
\end{equation}
 and the Hamiltonian is invariant under the TR transformation of  
$\Theta'$, $[H,\Theta']=0$. 
By rearranging the basis, $\gamma'_a$ (\ref{gammadash}) are transformed to the previous $SO(3,2)$ gamma matrices  $\gamma_a$ (\ref{so32gammamain}), and $\Theta'$ (\ref{thetadash}) is also to $\Theta$ (\ref{so32thetafirst}).   Thus, we have shown that the $J=3/2$ $SU(1,1)$ quadrupole Hamiltonian is equivalent to the $SO(3,2)$ Hamiltonian. 
 
\section{Summary and discussion}\label{sectsummary}

We explored the generalized Kramers degeneracy for $\Theta^2=+1$ in pseudo-hermitian quantum mechanics. 
As the quaternions realize the TR operation for $\Theta^2=-1$, the split-quaternions the TR operation for $\Theta^2=+1$.  
We showed, by passing from the quaternions to split-quaternions, the following generalized theorems in pseudo-hermitian quantum mechanics: 
\begin{itemize}
\item If the system is invariant under the TR transformation $\Theta^2=+1$ and also TR operator $\Theta$ is anticommutative with the metric operator, the system has at least doubly degenerate states: the generalized Kramers pair.  
\item  When the system is invariant under the particle/hole transformation $\mathcal{C}^2=+1$ and also charge-conjugation operator $\mathcal{C}$ is anticommutative with the metric operator, the system has paired states with $E$ and $-E$: the particle/hole pair.  
\end{itemize}
In both cases, the Hamiltonians necessarily 
possess the split-quaternion structure.  
 We also identified  TR, particle/hole, and pseudo-(anti-)hermitian symmetries in the non-hermitian category proposed by Bernard and LeClair \cite{Bernard2001}, and reconsidered the above theorems in view of non-hermitian random matrix.  
As a concrete example of the second theorem stated above, we 
investigated the $SU(1,1)$ model, and confirmed that the theorem indeed holds. 
Similarly, as an example of the first theorem, we introduced the $SO(3,2)$ model. We confirmed that     
the $SO(3,2)$  Hamiltonian is invariant under  TR transformation, and the TR symmetry  brings double degeneracy to the $SO(3,2)$  model, exactly analogous to the Kramers degeneracy of the $SO(5)$ model.  
The correspondences between the $SO(3,2)$ and  $J=3/2$ $SU(1,1)$  models are also clarified. 

 As pointed out in Ref.\cite{Avron1988}, the structure of the original $SO(5)$    model is related to instantons and twistor theory. Similarly, the present $SO(3,2)$    model is related to split-instantons \cite{Mason2006} and twistor theory \cite{HasebeTwistor2010}.  
The present work was inspired by recent developments of  the pseudo-hermitian quantum mechanics and the topological insulator in condensed matter physics. This work may hopefully be regarded as the first step of interplay between these two developments. It is intriguing to speculate realizations of  the pseudo-hermitian quantum mechanics in condensed matter physics.  
We would like to pursue the issue in a future research. 

\section*{Acknowledgements}

We acknowledge ISSP visiting program that enabled the collaboration. 
K.E.  was supported in part by Global COE Program
``the Physical Sciences Frontier,'' MEXT, Japan. 
This work was supported in part by a Grant-in Aid for Scientific
Research from MEXT of Japan, 
Grants No.22103005 (Innovative Areas ``Topological Quantum Phenomena''),
No.22540383, and No.23740212. 

\appendix 

\section{$SU(1,1)$ and $SO(3,2)$}
Here we briefly review the non-compact groups, $SU(1,1)$ and $SO(3,2)$.

The $SU(1,1)$ group consists of $2\times 2$ matrices $g$ satisfying the
following relations,
\begin{eqnarray}
g^{\dagger}\sigma_z g=\sigma_z,
\quad
{\rm det}g=1.
\end{eqnarray} 
Expanding $g$ by its generators, $\tau_i$ ($i=x,y,z$), $g=
1+i\sum_i\theta_i\tau_i+\cdots$ with real parameters $\theta_i$, we obtain 
\begin{eqnarray}
\tau_i^{\dagger}\sigma_z-\sigma_z\tau_i=0,
\quad
{\rm tr}\tau_i=0. 
\end{eqnarray} 
Thus the generators of $SU(1,1)$ are given by
\begin{eqnarray}
\tau_x=i\sigma_x,
\quad
\tau_y=i\sigma_y,
\quad
\tau_z=\sigma_z, 
\label{SU(1,1)matrices}
\end{eqnarray}
where $\sigma_i$ ($i=x,y,z$) are the standard Pauli matrices
for $SU(2)$ group. 
The $SU(1,1)$ Pauli matrices,  
$\tau_i=(\tau_x,\tau_y,\tau_z)=(i\sigma_x,i\sigma_y,\sigma_z)$, 
satisfy the following relations:  
\begin{subequations}
\begin{align}
&\sigma_x \tau_i{\sigma_x}^{-1}=-\tau_i^*,\\
&\sigma_y \tau_i {\sigma_y}^{-1}= -{\tau_i}^t,\\
&\sigma_z \tau_i {\sigma_z}^{-1}={\tau_i}^{\dagger}.
\end{align}
\end{subequations}

The $SO(3,2)$ group is linear transformations with unit determinant acting on a
five dimensional vector $(x_1,x_2,x_3,x_4,x_5)$ and preserving the
following norm,
\begin{eqnarray}
-x_1^2-x_2^2+x_3^2+x_4^2+x_5^2. 
\end{eqnarray}
Its element is given by a $5\times 5$
matrix $G$ which satisfies,
\begin{eqnarray}
\sum_{cd}G_{ac}G_{bd}\eta_{cd}=\eta_{ab},
\quad 
{\rm det}G=1, 
\end{eqnarray}  
with $\eta_{ab}={\rm diag}(-1,-1,1,1,1)$.   
Expanding $G$ by its generators $M^{ab}$
$(a,b=1,2,\cdots, 5)$,  
we find 
\begin{eqnarray}
(M^{ab})_{ce}\eta_{ed}+(M^{ab})_{de}\eta_{ce}=0, 
\quad
(M^{ab})_{cc}=0,
\end{eqnarray}
where we use the convention in which the repeating indices are summed.
These relations are met by the following $M^{ab}$,
\begin{eqnarray}
(M^{ab})_{cd}=i(\delta_{ca}\eta_{bd}-\delta_{cb}\eta_{ad}), 
\end{eqnarray}
which satisfies
\begin{eqnarray}
[M^{ab}, M^{cd}]
=-i(\eta_{ac}M^{bd}-\eta_{bc}M^{ad}-\eta_{ad}M^{bc}-\eta_{bd}M^{ac}). 
\label{eq:so32algebra}
\end{eqnarray}

Now consider the spinor representation of $SO(3,2)$. The spinor representation 
is given by the $4\times 4$ matrices $\gamma_a$ with anticommutation relations
\begin{eqnarray}
\{\gamma_a, \gamma_b \}=2\eta_{ab}.
\end{eqnarray}
The anticommutation relation is obtained by
\begin{eqnarray}
&&\gamma_1=
\left(\begin{array}{cc}
0 & \sigma_x \\
-\sigma_x & 0 
\end{array}\right),\gamma_2=
\left(\begin{array}{cc}
0 & \sigma_y\\
-\sigma_y & 0 
\end{array}\right),\gamma_3=
\left(\begin{array}{cc}
0 & -i\sigma_z \\
i\sigma_z & 0 
\end{array}\right),
\nonumber\\
&&\gamma_4=
\left(\begin{array}{cc}
0 &  1_2 \\
1_2 & 0 
\end{array}\right),\gamma_5=
\left(\begin{array}{cc}
1_2 & 0 \\
0 & -1_2 
\end{array}\right),  
\label{eq:gammaa}
\end{eqnarray}
and the $SO(3,2)$ algebra (\ref{eq:so32algebra}) is realized by 
\begin{eqnarray}
M^{ab}=\gamma_{ab}\equiv \frac{1}{4i}[\gamma_a,\gamma_b]. 
\label{eq:gammaab}
\end{eqnarray}

The gamma matrices (\ref{eq:gammaa}) and  generators
(\ref{eq:gammaab}) of $SO(3,2)$ satisfy the following relations, 
\begin{subequations}
\begin{align}
&k\gamma_a k^{-1}={\gamma_a}^*,\quad \quad\quad k\gamma_{ab} k^{-1}=-{\gamma_{ab}}^*,\\
&c\gamma_a c^{-1}={\gamma_a}^t,\quad\quad\quad c\gamma_{ab} c^{-1}=-{\gamma_{ab}}^t,
\label{chargeso32conj}\\
&{q} \gamma_a {q}^{-1} ={\gamma_a}^{\dagger}, \quad \quad\quad {q} \gamma_{ab} {q}^{-1} ={\gamma_{ab}}^{\dagger}, 
\label{hermitiangamma32}
\end{align}
\end{subequations}
where 
\begin{equation}
k=\begin{pmatrix}
\sigma_x & 0 \\
0 & \sigma_x
\end{pmatrix},~~~~
c=
\begin{pmatrix}
\sigma_y & 0 \\
0 & \sigma_y 
\end{pmatrix},~~~~
q=
\begin{pmatrix}
\sigma_z & 0\\
0 & \sigma_z
\end{pmatrix}. 
\end{equation}

\section{Quaternion and split quaternion}\label{appendquaternionsplit}
The quaternion $(1,e_1,e_2,e_3)$ are defined so as to satisfy \cite{Hamilton1844}
\begin{eqnarray}
&(e_1)^2=-1, ~~(e_2)^2=-1, ~~(e_3)^2=-1,\nonumber\\
&e_ie_j=-e_je_i ~~(i\neq j),~~~e_1e_2e_3=-1.
\label{quaternionsalgebra}
\end{eqnarray}
The ``imaginary'' quaternions are realized as Pauli matrices as   
\begin{equation}
(e_1,e_2,e_3)= (-i\sigma_x, -i\sigma_y, -i\sigma_z). 
\label{pauliquaternions}
\end{equation}
The split-quaternion algebra $(1,q_1,q_2,q_3)$ is simply obtained by flipping two signs of squares of quaternions:
\begin{eqnarray}
&(q_1)^2=+1, ~(q_2)^2=+1,~ (q_3)^2=-1,\nonumber\\
&q_iq_j=-q_jq_i ~~(i\neq j),~~~q_1q_2q_3=-1. 
\label{splitquaternionsalgebra}
\end{eqnarray}
The split-quaternions are realized by non-hermitian matrices as 
\begin{equation}
(q_1,q_2,q_3)= (i\tau_x,i\tau_y,i\tau_z)=(-\sigma_x,-\sigma_y,i\sigma_z),  
\label{idetificationsplitandtau}
\end{equation}
where $\tau_i$ $(i=x,y,z)$ denote the $SU(1,1)$ ``Pauli matrices'' 
(\ref{SU(1,1)matrices}). They satisfy 
\begin{equation}
 [\tau_i,\tau_j]=2i\epsilon_{ijk}\tau^k, \quad \quad\{\tau_i,\tau_j\}=-2\eta_{ij},
\label{su11splitalgebra}
\end{equation}
where $\epsilon_{ijk}$ is the three rank antisymmetric tensor 
with $\epsilon_{xyz}=1$, while $\eta_{ij}=diag(+1,+1,-1)$ and 
$\tau^i=(\tau_x,\tau_y,-\tau_z)$. 
Replacing the imaginary unit $i$ in $\sigma_y$ (\ref{pauliquaternions}) with three imaginary quaternions, the Pauli matrices are ``enhanced'' to yield $SO(5)$ gamma matrices:  
\begin{equation}
\left(\begin{array}{cc}
0 & i\sigma_x \\
-i\sigma_x & 0 
\end{array}\right),~~
\left(\begin{array}{cc}
0 & i\sigma_y \\
-i\sigma_y & 0 
\end{array}\right),~~
\left(\begin{array}{cc}
0 & i\sigma_z \\
-i\sigma_z & 0 
\end{array}\right),~~
\left(\begin{array}{cc}
0 &  1_2 \\
1_2 & 0 
\end{array}\right),~~
\left(\begin{array}{cc}
1_2 & 0 \\
0 & -1_2 
\end{array}\right).
\label{so5gammamatr}
\end{equation} 
 It is straightforward to see that (\ref{so5gammamatr}) satisfy  $\{\gamma_a,\gamma_b\}=2\delta_{ab}$. 
By applying such substitution in the case of split-quaternions, 
we obtain $4\times 4$ non-hermitian gamma matrices of $SO(3,2)$ (\ref{eq:gammaa}). 
The correspondence can also be naturally understood by noticing  isomorphism of groups: 
$SU(2)\simeq USp(2)$ and  $SO(5)\simeq USp(4)$;   
$SU(1,1)\simeq Sp(2,R)$ and $SO(3,2)\simeq Sp(4,R)$.  

\section{Definitions and relations for split-quaternions}\label{appendsplit}

 We introduce the terminology for split-quaternions  in the same spirit 
for quaternions (see Refs.\cite{Dyson1962, MehtaBook} for instance).  
The split-quaternion generally takes the form of 
\begin{equation}
q=cq_0 +c_iq_i,  
\label{compsplitquat}
\end{equation}
where $c$ and $c_i$ $(i=1,2,3)$ are complex numbers. There are three types of conjugation for split-quaternion: 
The complex conjugate, split-quaternionic conjugate, and split-quaternionic hermitian conjugate, which are respectively defined by  
\begin{subequations}
\begin{align}
&{q}^*=c^*q_0 +c_i^*q_i, \label{complexsplit} \\ 
&\overline{{q}}=cq_0 -c_iq_i, \label{defhermitiansplit} \\
&{q}^{\ddagger}\equiv {{\overline{({q}^*})}}=c^*q_0 -c_i^*q_i. 
\label{defsplitquaternionic} 
\end{align}\label{complexdefinitions}
\end{subequations}
Such conjugations have the following properties: 
\begin{equation}
(q_1\cdot q_2)^*={q_1}^*\cdot {q_2}^*,   \quad \quad \overline{(q_1 \cdot q_2)}=\overline{q_2}\cdot \overline{q_1}, \quad\quad (q_1\cdot  q_2)^{\ddagger}={q_2}^\ddagger \cdot {q_1}^\ddagger. 
\end{equation}
With the matrix realization (\ref{idetificationsplitandtau}), split-quaternion (\ref{compsplitquat}) is expressed as 
\begin{equation}
q= c+c_i i\tau_i= c-c_1 \sigma_x-c_2\sigma_y+c_3 i\sigma_z= \left( 
\begin{array}{cc}
c+ic_3 & -c_1+ic_2 \\
-c_1-ic_2 & c-ic_3
\end{array}
\right).  
\end{equation}
Correspondingly, the three kinds of conjugate (\ref{complexdefinitions}) are 
\begin{subequations}
\begin{align}
&{q}^{*}= c^*+c_i^* i\tau_i= c^*-c_1^* \sigma_x-c_2^*\sigma_y+c_3^* i\sigma_z   = \left( 
\begin{array}{cc}
c^*+ic_3^* & -c_1^*+ic_2^* \\
-c_1^*-ic_2^* & c^*-ic_3^*
\end{array}
\right),\\
&\overline{{q}}= c-c_i i\tau_i= c+c_1 \sigma_x+c_2\sigma_y-c_3 i\sigma_z   = \left( 
\begin{array}{cc}
c-ic_3 & c_1-ic_2 \\
c_1+ic_2 & c+ic_3
\end{array}
\right),\\
&{q}^{\ddagger}= c^*-c_i^* i\tau_i= c^*+c_1^* \sigma_x+c_2^*\sigma_y-c_3^* i\sigma_z   = \left( 
\begin{array}{cc}
c^*-ic_3^* & c_1^*-ic_2^* \\
c_1^*+ic_2^* & c^*+ic_3^*
\end{array}
\right). \label{splitmatsplithermiconj}
\end{align}
\end{subequations}
Due to the non-hermitian property of the split-quaternions, ${\tau_i}^{\dagger}=(-\tau_x,-\tau_y,\tau_z)$, the split-quaternionic hermitian conjugate (\ref{splitmatsplithermiconj}) does $\it{not}$ coincide with the ordinary definition of the hermitian conjugate 
\begin{equation}
{q}^{\dagger}=  c^*-c_i^* i{\tau_i}^{\dagger}= c^*-c_1^* \sigma_x-c_2^*\sigma_y-c_3^* i\sigma_z  = 
\left( 
\begin{array}{cc}
c^*-ic_3^* & -c_1^*+ic_2^* \\
-c_1^*-ic_2^* & c^*+ic_3^*
\end{array}
\right).   
\end{equation} 
(In the quaternion case, the quaternionic hermitian conjugate coincides with the ordinary hermitian conjugate.) 

The real split-quaternion is defined as
\begin{eqnarray}
{q}^r=wq_0 +x^iq_i=
\left(\begin{array}{cc}
w+ix^3 & -x^1+ix^2 \\
-x^1-ix^2 & w-ix^3
\end{array}
\right),  
\label{realquaterniondef}
\end{eqnarray}
 where $w$ and $x^i$ $(i=1,2,3)$ are real numbers. The necessary and sufficient condition for the real split-quaternion is given by 
\begin{equation}
 {q}^{\ddagger}=\overline{q}. 
\end{equation}
 An $M\times M$ split-quaternion matrix ($2M\times 2M$ matrix in the usual sense) Q is defined as a matrix whose matrix elements are split-quaternions: 
\begin{equation}
 (Q)_{IJ}=q_{IJ}, 
\end{equation}
where $I,J=1,2,\cdots,M$.  
The complex conjugation, split-quaternionic conjugation, and  split-quaternionic hermitian conjugation of ${Q}$, are respectively  defined as 
\begin{subequations}
\begin{align}
&({Q}^{*})_{IJ}={q_{IJ}}^{*}, \\
&(\overline{Q})_{IJ}=\overline{q_{JI}}, \\   
&({Q}^{\ddagger})_{IJ}={q_{JI}}^{\ddagger}. 
\end{align}
\end{subequations}
We call $\overline{Q}$ the ``dual'' of $Q$. The split-quaternionic hermitian matrix is a split-quaternion matrix that satisfies  
\begin{equation}
 {Q}^{\ddagger}=Q. 
 \label{splitquacond}
\end{equation}
 Unlike quaternion matrix, the split-quaternionic hermitian matrix is $\it{not}$ hermitian in the usual sense. 
 For instance, $q=w+ix^i q_i=w-x^i\tau_i=\begin{pmatrix} w-x^3 & -ix^1-x^2 \\-ix^1+x^2 & w+x^3 \end{pmatrix}$  (with real numbers $w,x^1,x^2,x^3$) is split-quaternionic hermitian, but not hermitian in the usual sense.   
 A real split-quaternionic matrix refers to the matrix whose components are real split-quaternions ${q}^r_{IJ}$, 
\begin{equation}
({Q}^r)_{IJ}={q}^r_{IJ}, 
\end{equation}
 and then ${Q}^r$ satisfies the relation 
\begin{equation}
 {Q}^{\ddagger}=\overline{Q}. 
 \label{realsplitqucond}
\end{equation}
 
 The self-dual real split-quaternion matrix is defined as a split-quaternion matrix that satisfies both (\ref{splitquacond}) and (\ref{realsplitqucond}): 
\begin{equation}
 Q={Q}^{\ddagger}=\overline{Q}. 
 \label{selfdualsplithermitiancond}
\end{equation}
Thus, the split-quaternionic hermitian real split-quaternion matrix is equivalent to the  
 self-dual real split-quaternion matrix.  (The condition $\overline{Q}=Q$ is the self-dual condition.)  The terminology ``split-quaternionic hermitian real split-quaternion'' is rather clumsy, so we use ``self-dual real split-quaternion'' instead.  
 Such  self-dual real split-quaternion matrix generally accommodates the  generalized Kramers degeneracy for $\Theta^2=+1$ \footnote{
Real split-quaternion matrix does not accommodate the generalized Kramers degeneracy in general. The split-quaternionic hermitian condition has to be imposed as well.}.   
 In low dimensions, the self-dual real split-quaternion matrices are given by 
 \begin{align}
&M=1\quad : \quad  Q_1 =\left(\begin{array}{cc} 
w & 0 \\
0 & w 
\end{array}\right)=w 1_2 ,  
 \nonumber\\
&M=2\quad : \quad  Q_2 =\left(\begin{array}{cccc} 
w+x^5 & 0 & x^4-ix^3 & x^1-ix^2 \\
0 & w+x^5 &   x^1+ix^2   & x^4+ix^3    \\
 x^4+ix^3     & -x^1+ix^2    & w-x^5 & 0 \\
 -x^1-ix^2    &  x^4-ix^3    & 0  & w-x^5
\end{array}\right),
\label{lowsplitselfdualmat}
 \end{align}
 where $w, x^1,\cdots,x^5$  are real parameters. 
With the $SO(3,2)$ gamma matrices (\ref{so32gammamain}), $Q_2$ is concisely represented as 
\begin{equation}
Q_2=w 1_4+\sum_{a=1}^5x^a\gamma_a.
\end{equation}
$Q_1$ and $Q_2$ are exactly equal to the matrices (\ref{2times2unitmat}) 
and (\ref{gammaso33hamil}), respectively.  They have both K and Q symmetries $(\epsilon_q=+1)$, and their eigenvalues are 
\begin{align}
&M=1: ~~~ w\nonumber\\
&M=2:  ~~~~w\pm \sqrt{-(x^1)^2-(x^2)^2+(x^3)^2+(x^4)^2+(x^5)^2},
\end{align}
with double degeneracy (of the generalized Kramers).  
 
\section{Spin flipping operators and quadrupole Hamiltonians for low $SU(1,1)$ spins}\label{appensu11spinquadrupole}
 
The $SU(1,1)$ algebra is given by 
\begin{equation}
[J_i,J_j]=i\epsilon_{ijk}J^k, 
\label{algebrasu11}
\end{equation}
where $\epsilon_{ijk}$ denotes a totally antisymmetric tensor 
with $\epsilon_{xyz}=1$, and $J^i=(J_x,J_y,-J_z)$. 
Explicitly. 
\begin{equation}
[J_x,J_y]=-iJ_z,~~[J_y,J_z]=iJ_x,~~[J_z,J_x]=iJ_y. 
\end{equation}
From  $SU(2)$ spins $S_x,S_y,S_z$, the $SU(1,1)$ spins are constructed with  
 the identification 
\begin{equation}
J_x=iS_x,~~J_y=iS_y,~~J_z=S_z. 
\end{equation}
 Note that $J_x$ is pure imaginary; $J_y$ and $J_z$ are real. 
The magnitude of $SU(1,1)$ spin $J$ is defined as 
\begin{equation}
-J_x^2-J_y^2+J_z^2=J(J+1). 
\end{equation}
For instance,
\begin{subequations}
\begin{align}
J=1/2~:~&J_x=\frac{1}{2}\tau_x,~~~~~J_y=\frac{1}{2}\tau_y,~~~~~J_z=\frac{1}{2}\tau_z,~~
\label{psedotimereversal12_2} \\
J=1~:~&
J_x=\frac{1}{\sqrt{2}}
\left(\begin{array}{ccc} 
0 & i  & 0 \\
i & 0 & i \\
0 & i & 0  
\end{array}\right), ~~
J_y=\frac{1}{\sqrt{2}}
\left(\begin{array} {ccc}
0 & 1  & 0  \\
-1 & 0 & 1  \\
0 & -1 & 0 
\end{array}\right),~~J_z=
\left(\begin{array}{ccc}
1 & 0 & 0  \\
0 & 0 & 0  \\
0 & 0 & -1 
\end{array}\right).
\end{align} 
\end{subequations}
The $SU(1,1)$ spin flipping operator $\Xi$ is defined so as to satisfy 
\begin{equation}
J_i~~\rightarrow ~~ \Xi J_i\Xi^{-1}=-J_i. 
\end{equation}
Express $\Xi=U\cdot K$, and $U$ satisfies  
\begin{equation}
U J_x U^{-1}=J_x,~~U J_y U^{-1}=-J_y,~~U J_z U^{-1}=-J_z.  
\end{equation}
Here, $U$ is a unitary matrix given by   
\begin{equation}
U=(-i)^{2J}e^{\pi J_x}, 
\end{equation}
where the factor $(-i)^{2J}$ is added for convenience. 
Consequently, $\Xi$ is given by 
\begin{equation}
\Xi=(-i)^{2J}e^{\pi J_x}\cdot K,    
\label{pseudoTRoperatorgeneral}
\end{equation}
which satisfies 
\begin{equation}
\Xi^2=+1,
\end{equation}
independent of the magnitude of the $SU(1,1)$ spin\footnote{ As discussed in Sec.\ref{subsec:trforboson}, the square of the $SU(2)$ spin flipping operator takes $-1$ for half-integer spins, while the $SU(1,1)$ spin flipping operator yields +1 even for half-integer $SU(1,1)$ spins.}. 
For low $SU(1,1)$ spins, 
\begin{subequations}
\begin{align}
J=1/2~:~&\Xi_{1/2}=\sigma_x\cdot K,
\label{psedotimereversal12_3} \\
J=1~~:~
&\Xi_{1}=
\left(\begin{array}{ccc}
 0 & 0 & 1  \\
0 & 1 & 0 \\
1 & 0 & 0
\end{array}\right)\cdot K, \\
J=3/2~:~
&\Xi_{3/2}=
\left(\begin{array}{cc}
 0 & \sigma_x  \\
\sigma_x & 0 
\end{array}\right)\cdot K. 
\label{tildethetaso32}
\end{align}
\end{subequations}
Note $\Xi_{1/2}$ is equal to the charge conjugation operator $\mathcal{C}$ 
(\ref{psedotimereversal12}) of the $SU(1,1)$ model, and $\Xi_{3/2}$ the time-reversal operator $\Theta'$ (\ref{thetadash}) of the $SO(3,2)$ model.  
Similarly, for low $SU(1,1)$ spins, the quadrupole Hamiltonians (\ref{quadrupolehamitlnians}) introduced in Sec.\ref{subsec:realizationasj=3/2} are given by  
\begin{subequations}
\begin{align}
J=1/2:~
&H(Q_a)=0, \\
J=1:~ 
&H(Q_{1})=\frac{1}{\sqrt{6}}
\left(\begin{array}{ccc}
0 & i & 0 \\
i & 0 & -i \\
0 & -i & 0 
\end{array}\right),
~~H(Q_2)=\frac{1}{\sqrt{6}}
\left(\begin{array}{ccc}
0 & 1 & 0 \\
-1 & 0 & -1 \\
0 & 1 & 0 
\end{array}\right), \nonumber\\
&H(Q_{3})=-\frac{1}{\sqrt{3}}
\left(\begin{array}{ccc}
0 & 0 & 1 \\
0 & 0 & 0 \\
1 & 0 & 0 
\end{array}\right),~~H(Q_4)=\frac{1}{\sqrt{3}}
\left(\begin{array}{ccc}
0 & 0 & i \\
0 & 0 & 0 \\
-i & 0 & 0 
\end{array}\right), ~~H(Q_{5})=\frac{1}{{3}}
\left(\begin{array}{ccc}
1 & 0 & 0 \\
0 & -2 & 0 \\
0 & 0 & 1 
\end{array}\right).
\end{align}
\end{subequations}

\section{${\cal P}$, ${\cal T}$, and ${\cal C}$ operators for $SU(1,1)$
 and $SO(3,2)$ models}
For non-hermitian Hamiltonians, there is a systematic procedure for
constructing a metric operator $\eta_0$, and symmetry operators which
commute with the Hamiltonian \cite{BBJ2002, Ahmed2003}.
In this Appendix, we briefly review the procedure and apply it to the
$SU(1,1)$ and $SO(3,2)$ models, respectively. 

For a non-hermitian Hamiltonian $H$, the Schr{\"o}dinger equation is
\begin{equation}
H|\phi_n\rangle =E_n |\phi_n\rangle,~~~~
H^{\dagger}|\varphi_n\rangle\!\rangle =E_n^* |\varphi_n\rangle\!\rangle,  
\label{scheq}
\end{equation}
where the eigenvalues $E_n$ are complex in general. As discussed in
Sec.\ref{generalizedKramers},
the eigenvectors, $|\phi_n\rangle$ and $|\varphi_n \rangle\!\rangle $, 
give a  bi-orthonormal basis \cite{Wong1967,FaisalandMoloney1981} satisfying the orthonormal and complete relations 
\begin{equation}
\langle\!\langle \varphi_m|\phi_n\rangle =\langle \phi_m|\varphi_n\rangle\!\rangle=\delta_{mn},
~~~~
\sum_n |\phi_n\rangle  \langle\!\langle  \varphi_n | =\sum_n |\varphi_n \rangle\!\rangle \langle  \phi_n|=1. 
\label{appendorthocompleterelation}
\end{equation}
The Hamiltonian and its hermitian conjugate are expanded as 
\begin{equation}
H=\sum_n E_n |\phi_n\rangle \langle \!  \langle \varphi_n|,~~~~H^{\dagger}=\sum_n E_n^*  |\varphi_n \rangle\!\rangle \langle  \phi_n|. 
\label{expansionsofhamiltonians}
\end{equation}
A non-hermitian Hamiltonian is called pseudo-hermite when it satisfies
\begin{equation}
H^{\dagger}=\eta H\eta^{-1},
\label{condpseudohermiticity}
\end{equation}
where $\eta$ is hermitian and called a metric operator.  
A state given by  
\begin{equation}
|\phi_n\rangle '\equiv \eta^{-1}|\varphi_n\rangle\!\rangle, 
\end{equation}
is an eigenvector of $H$ with an eigenvalue $E_n^*$ as seen from  (\ref{scheq}) and (\ref{condpseudohermiticity}).
Thus the eigenvalues of the pseudo-hermitian Hamiltonian are classified into two types. One is a set of real eigenvalues  and 
the other is a set of complex conjugate pairs. (For more details, see \cite{Mostafazadeh2008}.)
  
Suppose all the eigenvalues of a non-hermitian Hamiltonian are real, 
$i.e.$ $E_n^*=E_n$.  
Following Ref.\cite{Ahmed2003}, a metric operator $\eta_0$ is
constructed as   
\begin{equation}
\eta_0^{-1}=\sum_{l=1}^N |\phi_{l}\rangle \langle
 \phi_l|,~~~\eta_0=\sum_{l=1}^N |\varphi_{l}\rangle\!\rangle
 \langle\!\langle\varphi_l|. 
\label{realizationeta2}
\end{equation}
Here note that the metric operator satisfying (\ref{condpseudohermiticity})
is not unique. 
$\eta_0$ in the above generally depends on parameters of the
Hamiltonian, while we may have a constant metric operator for some models. 
For example, see Sec.\ref{subsec:su11} and Sec.\ref{subsec:so32}.

Let us now define the following operators $\mathcal{P}$, $\mathcal{T}$,
and $\mathcal{C}$,
\begin{subequations}
\begin{align}
&\mathcal{P}|\varphi_n \rangle\!\rangle =(-1)^{n+1}|\phi_n\rangle,\\ 
&\mathcal{T}|\phi_n\rangle=|\varphi_n \rangle\!\rangle,\\
&\mathcal{C}|\phi_n\rangle=(-1)^{n+1}|\phi_n\rangle. 
\end{align}
\end{subequations}
These operators were originally introduced in analogy with parity, TR, and charge conjugation, respectively,  however they are not, 
in fact, directly related. With the orthonormal and complete conditions (\ref{appendorthocompleterelation}), they are explicitly written as
\begin{subequations}
\begin{align}
&\mathcal{P}=\sum_{l=1}^N (-1)^{l+1} |\phi_l\rangle\langle\phi_l|,
\label{definitionofP}\\
&\mathcal{T}=\sum_{l=1}^N |\varphi_l\rangle\!\rangle  K\langle\!\langle\varphi_l|, 
\label{definitionofT}\\
&\mathcal{C}=\sum_{l=1}^N (-1)^{l+1} |\phi_l\rangle\langle\!\langle \varphi_l|,  
\end{align}\label{PTCoperatorsstate}
\end{subequations}
which are respectively hermitian, anti-hermitian\footnote{The terminology ``anti-hermitian'' usually refers to  operator whose hermitian conjugate is equal to the minus of the original, but here, the terminology ``anti-hermitian''  refers to  hermitian operator that anticommutes with imaginary unit.}, and pseudo-hermitian
\begin{subequations}
\begin{align}
&\mathcal{P}^{\dagger}=\mathcal{P},\\
&\mathcal{T}^{\dagger}=\mathcal{T},\\
&\mathcal{C}^{\dagger}=\eta_0~ \mathcal{C}~ \eta_0^{-1}. 
\end{align}
\end{subequations}
From (\ref{PTCoperatorsstate}), 
$\mathcal{PT}$ and $\mathcal{CPT}$ operators are given by 
\begin{subequations}
\begin{align}
&\mathcal{PT}=\sum_{l=1}^N (-1)^{l+1} |\phi_l\rangle  K \langle\!\langle \varphi_l|,  
\\
&\mathcal{CPT}=\sum_{l=1}^N |\phi_l\rangle K \langle\!\langle \varphi_l|,
\end{align}\label{PTCPToperatorsformulas}
\end{subequations}
and
\begin{subequations}
\begin{align}
&\mathcal{PT} |\phi_n\rangle=(-1)^{n+1} |\phi_n\rangle,\label{stateexpressPT}\\
&\mathcal{CPT}|\phi_n\rangle = |\phi_n\rangle.
\end{align}
\end{subequations}
They are pseudo-antiunitary  
\begin{subequations}
\begin{align}
&(\mathcal{PT})^{\dagger}=\eta_0 (\mathcal{PT})^{-1} \eta_0^{-1},
\label{pseudohermitePT}\\
&(\mathcal{CPT})^{\dagger}=\eta_0 (\mathcal{CPT})^{-1} \eta_0^{-1}. 
\end{align}
\end{subequations}
It is readily seen that these operators satisfy  
\begin{align}
&[H,\mathcal{PT}]=0,\nonumber\\
&[H,\mathcal{C}]=[H,\mathcal{CPT}]=0,\nonumber\\
&[\mathcal{PT},\mathcal{C}]=0,\nonumber\\
&[H,\mathcal{P}]\neq 0, 
\label{commutationrelationPTH}
\end{align}
and 
\begin{align}
&(\mathcal{PT})^2=\mathcal{C}^2=(\mathcal{CPT})^2=1,\nonumber\\
&\mathcal{P}^2\neq 1,~~~~\mathcal{T}^2\neq 1.
\end{align}
As realized in the first and the second lines of
(\ref{commutationrelationPTH}), the Hamiltonian always displays
``$\mathcal{PT}$ symmetry'' and ``${\cal C}$ symmetry '' 
(or ``${\cal CPT}$ symmetry '') with respect to the $\mathcal{PT}$
and ${\cal C}$ operators constructed above.  

\subsection{$SU(1,1)$ model}
\label{subsec:su11}

With the bi-orthonormal bases (\ref{biortho1phi}) and (\ref{biortho2varphi}), 
we construct the metric operator $\eta_0$, and ${\cal P}$, ${\cal T}$, and
${\cal C}$ operators for the $SU(1,1)$ model.  
From (\ref{realizationeta2}), 
\begin{equation}
\eta_0=|\varphi_+\rangle\!\rangle\langle\!\langle\varphi_+|
+|\varphi_-\rangle\!\rangle\langle\!\langle\varphi_-|=\frac{1}{r}
\begin{pmatrix}
z & -ix-y \\
ix -y & z 
\end{pmatrix},  
\label{eta12metric}
\end{equation}
 from (\ref{PTCoperatorsstate}), 
\begin{align}
&\mathcal{P}=|\phi_+\rangle\langle\phi_+|-|\phi_-\rangle\langle\phi_-|
=\sigma_z,\nonumber\\
&\mathcal{T}=(|\varphi_+\rangle\!\rangle\langle\!\langle\varphi_+^*|
+|\varphi_-\rangle\!\rangle\langle\!\langle\varphi_-^*|)\cdot K
=\frac{1}{2r(r+z)}\left(
\begin{array}{cc}
(r+z)^2+(ix+y)^2 & -2(r+z)y\\
 -2(r+z)y &  (r+z)^2+(ix+y)^2
\end{array}\right)\cdot K,\nonumber\\
&\mathcal{C}=|\phi_+\rangle\langle\!\langle\varphi_+|
-|\phi_-\rangle\langle\!\langle\varphi_-|=\frac{1}{r}
\left(\begin{array}{cc}
z & -ix-y \\
-ix+y & -z
\end{array}
\right)=\frac{1}{r}H, 
\end{align}
with $H$ (\ref{SU11H}), 
and from (\ref{PTCPToperatorsformulas}),   
\begin{align}
&\mathcal{PT}=(|\phi_+\rangle\langle\!\langle\varphi_+^*|
-|\phi_-\rangle\langle\!\langle\varphi_-^*|)\cdot K =
\frac{1}{2r(r+z)}
\left(
\begin{array}{cc} 
 (r+z)^2 +(ix+y)^2 & -2(r+z)y \\
 2(r+z)y  &  -(r+z)^2 -(ix-y)^2
 \end{array}\right)\cdot K,\nonumber\\
&\mathcal{CPT}= (|\phi_+\rangle\langle\!\langle\varphi_+^*|
+|\phi_-\rangle\langle\!\langle\varphi_-^*|)\cdot K = \frac{1}{2r(r+z)}
\left(
\begin{array}{cc} 
 (r+z)^2 -(ix+y)^2 & 2i(r+z)x \\
 -2i(r+z)x  &  (r+z)^2 -(ix-y)^2
 \end{array}\right)\cdot K. 
\end{align}
The metric operator (\ref{eta12metric}) is different from
$\eta=\sigma_z$ used in Sec.\ref{randommatrixcategorysu11}; This
``discrepancy'' stems from the non-uniqueness of the metric operator. 
 
\subsection{$SO(3,2)$ model}
\label{subsec:so32}
 
For $SO(3,2)$ model, with use of the bi-orthonormal basis 
$|\psi_{m\alpha}\rangle$ (\ref{biortho1psi}) and  
$|\chi_{m\alpha}\rangle\!\rangle$ (\ref{biortho2chi}), 
the metric operator is constructed as  
\begin{align}
\eta&_0=\sum_{\alpha=+,-}
(|\chi_{+\alpha}\rangle\!\rangle\langle\!\langle\chi_{+\alpha}|
+|\chi_{-\alpha}\rangle\!\rangle\langle\!\langle\chi_{-\alpha}|)
\nonumber\\
&=\frac{1}{2r(r+x^5)}\left(
\begin{array}{cc}
(r+x^5)^2-(x^4-ix^i\tau_i^{\dagger})(x^4+ix^j\tau_j) & -i(r+x^5)x^i(\tau_i-\tau_i^{\dagger}) \\
-i(r+x^5)x^i(\tau_i-\tau_i^{\dagger})& (r+x^5)^2+(x^4+ix^i\tau_i^{\dagger})(x^4-ix^j\tau_j)
\end{array}\right). 
\label{pseudometric}
\end{align}
Similar to the case of $SU(1,1)$ model,  
 the metric operator (\ref{pseudometric})  is different from the one given by  (\ref{so32q}).  (\ref{so32q}) is anticommutative with $\Theta$ (\ref{so32thetafirst}), while (\ref{pseudometric}) is commutative with $\Theta$.   
From (\ref{PTCoperatorsstate}), 
 $\mathcal{P}$, $\mathcal{T}$ and $\mathcal{C}$ are derived as 
\begin{align}
\mathcal{P}&=\sum_{\alpha=+,-}
(|\psi_{+\alpha}\rangle\langle\psi_{+\alpha}|-
|\psi_{-\alpha}\rangle\langle\psi_{-\alpha}|)
\nonumber\\&=\frac{1}{2r(r+x^5)}
\left(\begin{array}{cc}
(r+x^5)^2-(x^4-ix^i\tau_i)(x^4+ix^j\tau_j^{\dagger}) & (r+x^5)(2x^4-ix^i(\tau_i+\tau_i^{\dagger})) \\
 (r+x^5)(2x^4+ix^i(\tau_i+\tau_i^{\dagger})) & -(r+x^5)^2+(x^4+ix^i\tau_i)(x^4-ix^j\tau_j^{\dagger}) 
\end{array}\right),\nonumber\\
\mathcal{T}&=\sum_{\alpha=+,-}
(|\chi_{+\alpha}\rangle\!\rangle\langle\!\langle\chi_{+\alpha}^*|
+|\chi_{-\alpha}\rangle\!\rangle\langle\!\langle\chi_{-\alpha}^*|)
\cdot K \nonumber\\&=\frac{1}{2r(r+x^5)}\left( 
\begin{array}{cc}
(r+x^5)^2  +(x^4-ix^i\tau_i^{\dagger})(x^4-ix^j\tau_j^{*}) & i(r+x^5)x^i(\tau_i^*+\tau_i^{\dagger})\\
i(r+x^5)x^i(\tau_i^*+\tau_i^{\dagger}) &   (r+x^5)^2  +(x^4+ix^i\tau_i^{\dagger})(x^4+ix^j\tau_j^{*})
\end{array}\right) \cdot K,  \nonumber\\
\mathcal{C}&= \sum_{\alpha=+,-}
(|\psi_{+\alpha}\rangle\langle\!\langle\chi_{+\alpha}|-
|\psi_{-\alpha}\rangle\langle\!\langle\chi_{-\alpha}|)\nonumber\\
&= \frac{1}{r}
\begin{pmatrix}
x^51_2 & x^4 1_2 -ix^i\tau_i \\
x^4+ix^i\tau_i & -x^51_2
\end{pmatrix}     
=\frac{1}{r}H, 
\end{align}
with $H$ (\ref{SO32hamiltonian}), 
and $\mathcal{PT}$ and $\mathcal{CPT}$ are 
\begin{align}
\mathcal{PT}&=\sum_{\alpha=+,-}
(|\psi_{+\alpha}\rangle\langle\!\langle\chi^*_{+\alpha}|
-|\psi_{-\alpha}\rangle\langle\!\langle\chi_{-\alpha}^*|)\cdot K\nonumber\\
&=
\frac{1}{2r(r+x^5)}\left( 
\begin{array}{cc}
 (r+x^5)^2 - (x^4-ix^i\tau_i)(x^4-ix^j\tau_j^*) & (r+x^5)(2x^4-ix^i(\tau_i-\tau_i^{*})) \\
(r+x^5)(2x^4+ix^i(\tau_i-\tau_i^{*}))& -(r+x^5)^2 + (x^4+ix^i\tau_i)(x^4+ix^j\tau_j^*)
\end{array}
\right)\cdot K, \nonumber\\
\mathcal{CPT}&=\sum_{\alpha=+,-}
(|\psi_{+\alpha}\rangle\langle\!\langle\chi_{+\alpha}^*|
+|\psi_{-\alpha}\rangle\langle\!\langle\chi_{-\alpha}^*|)
\cdot K\nonumber\\
&=
\frac{1}{2r(r+x^5)}\left( 
\begin{array}{cc}
 (r+x^5)^2 + (x^4-ix^i\tau_i)(x^4-ix^j\tau_j^*) & i(r+x^5)x^i(\tau_i+\tau_i^{*})) \\
i(r+x^5)x^i(\tau_i+\tau_i^{*}) & (r+x^5)^2 + (x^4+ix^i\tau_i)(x^4+ix^j\tau_j^*) 
\end{array}
\right)\cdot K. 
\end{align}
 
\section{Level crossing point and monopole structure in non-hermitian
 Hamiltonians} 

\subsection{$SU(1,1)$ model and $U(1)$ monopole}

Singularity of phase of eigenstate generally reflects the non-trivial topology in phase space \cite{TKNN1982,Kohmoto1985}.  
For instance, 
the crossing point of two energy levels of the $SU(2)$ model is called a diabolic point (an isolated point), which brings $U(1)$ holonomy in phase space \cite{Berry1984}. 
In the $SU(1,1)$ model, the eigen-energies are given by $E_{\pm}=\pm r$ with $r=\sqrt{z^2-x^2-y^2}$, and the level crossing point $E_+=E_-$ is achieved when $r=0$. 
 In the $SU(2)$ model, because of the Euclidean signature, the condition, $r=\sqrt{x^2+y^2+z^2}=0$, is met only at the point $x=y=z=0$. Meanwhile, in the $SU(1,1)$ model, the signature is hyperbolic, and  the condition is satisfied on the surface 
\begin{equation}
x^2+y^2-z^2=0. 
\end{equation}
In such a case, $r=0$ point is called the exceptional point.  
Around the exceptional point ($r\sim 0$), the upper and lower energy eigenvectors $(z\ge r)$ (\ref{biortho1phi}) behave as   
\begin{equation}
|\phi_+\rangle \sim  \frac{1}{\sqrt{2rz}}
\left(\begin{array}{c}
z\\
y-ix
\end{array}\right),~~~  |\phi_-\rangle \sim \frac{1}{\sqrt{2rz}}
\left(\begin{array}{c}
y+ix\\
z
\end{array}\right), 
\end{equation}
and 
\begin{equation}
|\varphi_+\rangle\!\rangle \sim \frac{1}{\sqrt{2rz}}
\left(\begin{array}{c}
z\\
-y+ix
\end{array}\right),~~~|\varphi_-\rangle\!\rangle \sim \frac{1}{\sqrt{2rz}}
\left(\begin{array}{c}
-y-ix\\
z
\end{array}\right).  
\end{equation}
They are degenerate, as found  
\begin{align}
&|\phi_-\rangle \sim e^{i\chi}|\phi_+\rangle,\nonumber\\
&|\varphi_-\rangle\!\rangle\sim -e^{i\chi}|\varphi_+\rangle\!\rangle, 
\end{align}
where $\tan\chi=\frac{x}{y}$. 
Then, the normalization condition is not satisfied but  
\begin{equation}
\langle \phi_+|\varphi_+\rangle\!\rangle = 
\langle\phi_-|\varphi_-\rangle\!\rangle
\sim  \frac{r}{2z}\sim 0, 
~~~~\langle\phi_+|\varphi_-\rangle\!\rangle= 0,~~~~ 
\langle\phi_-|\varphi_+\rangle\!\rangle= 0. 
\end{equation}

The holonomy of the exceptional points is 
\begin{equation}
A^{+}=-i\langle\!\langle\varphi_+|d\phi_+\rangle
=  -i\langle\phi_+|d\varphi_+\rangle\!\rangle  
=-i\langle\phi_{+}|\sigma_z |d\phi_+\rangle=dx^i A^{+}_i, 
\end{equation}
where 
\begin{equation}
A^{+}_i=\frac{1}{2r(r+z)}\epsilon_{ij3}x^j,
\end{equation}
which is the $U(1)$ monopole gauge field in a hyperbolic space \cite{Nesterov2008,Hasebe2009}.  
The corresponding field strength is calculated as 
\begin{equation}
F_{ij}^{+}=\partial_i A^{+}_j-\partial_j A^{+}_i
=\frac{1}{2r^3}\epsilon_{ijk}x^k.
\end{equation}
Similarly, the holonomy for the negative energy state is evaluated as  
\begin{equation}
A^{-}=-i\langle\!\langle\varphi_-|d\phi_-\rangle
=  -i\langle\phi_-|d\varphi_-\rangle\!\rangle
=-i\langle\phi_{-}|(-\sigma_z)|d\phi_-\rangle=dx^i A_i^{-}, 
\end{equation}
where the insertion of $-\sigma_z$ is in accordance with the normalization (\ref{normalizationforphipm}), and 
\begin{equation}
A_i^{-}=-\frac{1}{2r(r+z)}\epsilon_{ij3}x^j=-A_i^+. 
\end{equation}
The corresponding gauge field strength is 
\begin{equation}
F_{ij}^{-}=\partial_i A_j^{-}-\partial_j A_i^{-}=-\frac{1}{2r^3}\epsilon_{ijk}x^k=-F_{ij}^+.
\end{equation}
The field strength diverges at the exceptional point.  

On the lower leaf $(z\le -r)$, the holonomies are derived as 
\begin{subequations}
\begin{align}
&A_+'=-i\langle\!\langle{\varphi'}_+|d\phi'_+\rangle
=  -i\langle{\phi'}_+|d\varphi'_+\rangle\!\rangle
=-i\langle{\phi'}_+|(-\sigma_z)|d\phi_+'\rangle=dx^i {A_i'}^{+},\\
&A_-'=-i\langle\!\langle{\varphi'}_-|d\phi'_- \rangle
=-i\langle{\phi'}_-|d\varphi'_-\rangle\!\rangle
=-i\langle{\phi'}_-|\sigma_z|d\phi_-'\rangle=dx^i {A_i'}^{-}, 
\end{align}
\end{subequations}
where
\begin{equation}
{A_i'}^{+}=-{A_i'}^{-}=-\frac{1}{2r(r-z)}\epsilon_{ij3}x^j.  
\end{equation}

As found above, the holonomy of the exceptional point in the $SU(1,1)$ model is regarded as the gauge field of hyperbolic $U(1)$ monopole, and the monopole charges for the upper and lower energy states are opposite.    
Such effect is similar to the $U(1)$ monopole holonomy of the diabolic point in the $SU(2)$ model \cite{Berry1984}. 

\subsection{$SO(3,2)$ model and $SU(1,1)$ monopole}

 Degeneracies in energy levels generally bring non-abelian holonomy  in
 the parameter space  \cite{Wilczek1984}.  
For instance, the $SO(5)$ Hamiltonian (Luttinger Hamiltonian \cite{Luttinger}) 
has $SU(2)$ holonomy which is crucial for the spin-Hall effect 
\cite{Murakamietal2003}. Here, we consider what kind of holonomy could emerge in the $SO(3,2)$ model.  
The energy levels of the $SO(3,2)$ model are $\pm r$, and
the level crossing point is at $r=0$, namely
\begin{equation}
(x^1)^2+(x^2)^2-(x^3)^2-(x^4)^2-(x^5)^2=0.
\label{exceptionalso32cond}
\end{equation}
Near the exceptional point $(r\sim 0)$, 
the upper energy and lower energy eigenvectors behave as  
\begin{equation}
|\psi_{+\alpha}\rangle  \sim\frac{1}{\sqrt{2rx^5}}
\left(\begin{array}{c}
x^5\phi_{\alpha}\\
(x^4+ix^i\tau_i)\phi_{\alpha}
\end{array}\right),~~~|\psi_{-\alpha}\rangle  \sim\frac{1}{\sqrt{2rx^5}}
\left(\begin{array}{c}
-(x^4-ix^i\tau_i)\phi_{\alpha}\\
x^5\phi_{\alpha}
\end{array}\right),
\end{equation}
and  
\begin{equation}
|\chi_{+\alpha}\rangle\!\rangle \sim  \frac{1}{\sqrt{2rx^5}}
\left(\begin{array}{c}
x^5\phi_{\alpha}\\
(x^4+ix^i\tau_i^{\dagger})\phi_{\alpha}
\end{array}\right),~~~
|\chi_{-\alpha}\rangle\!\rangle \sim \frac{1}{\sqrt{2rx^5}}
\left(\begin{array}{c}
-(x^4-ix^i\tau_i^{\dagger})\phi_{\alpha}\\
x^5\phi_{\alpha}
\end{array}\right).
\end{equation}
Then, at the exceptional point, 
$|\psi_{+\alpha}\rangle$ and $|\psi_{-\alpha}\rangle$  are related by the $SU(1,1)$ gauge transformation $-\frac{x^4-ix^i\tau_i}{x^5}$, and 
$|\chi_{+\alpha}\rangle\!\rangle$ and $|\chi_{-\alpha}\rangle\!\rangle$
are by $-\frac{x^4-ix^i\tau_i^{\dagger}}{x^5}$,  and the normalization
conditions are no longer satisfied:
\begin{equation}
\langle\psi_{\pm \alpha}|\chi_{\pm\alpha'}\rangle\!\rangle
\overset{r\sim 0}\sim \frac{r}{2x^5}\sim 0,~~~\langle\psi_{\pm \alpha}|\chi_{\mp\alpha'}\rangle\!\rangle= 0.
\end{equation}

For $x^5 > -r$, the upper energy degenerate eigenvectors bring the following holonomy:  
\begin{equation}
A^+=-i\langle\!\langle\chi_+|d\psi_{+}\rangle
=-i\langle\psi_+|d\chi_+\rangle\!\rangle =dx^a \phi^{\dagger}\sigma_z A_a^+ \phi
\end{equation}
where 
\begin{equation}
A^+_{\mu}=-  \frac{1}{2r(r+x^5)}\eta_{\mu\nu i} x^{\nu}\tau^i,~~~~A^+_5=0. 
\end{equation}
Here, $\mu,\nu=1,2,3,4$, and $\eta_{\mu\nu i}$ are the ``split''-'t Hooft symbol defined by $\eta_{\mu\nu i}=\epsilon_{\mu\nu i}+\eta_{\mu i}\eta_{\nu 4}-\eta_{\nu i}\eta_{\mu 4}$ with 
$\eta_{\mu\nu}=diag(+,+,-,-)$, and $\epsilon_{\mu\nu i}$ is a totally antisymmetric tensor, $\epsilon_{123}=1$ and $\epsilon_{\mu\nu i}=0$ for $\mu=4$ or $\nu=4$.  
The corresponding field strength 
\begin{equation}
F_{ab}^+=\partial_a A^+_b-\partial_b A^+_a +i[A^+_a,A^+_b],
\end{equation}
is derived as  
\begin{align}
&F^+_{\mu\nu}=\frac{1}{r^2}x_{\mu}A^+_{\nu}-\frac{1}{r^2}x_{\nu}A^+_{\mu}-\frac{1}{2r^2}\eta_{\mu\nu i}\tau^i,\nonumber\\
&F^+_{\mu5}=-F^+_{5\mu}=\frac{1}{r^2}(r+x^5)A^+_{\mu}. 
\end{align}
The gauge field strength has the singularity at the exceptional point. 
Similarly, $A^-$ is given by 
\begin{equation}
{A}^-=-i\langle\!\langle{\chi}_-|d\psi_-\rangle
=-i\langle{\psi_-}|d\chi_-\rangle\!\rangle 
=dx^a {\phi}^{\dagger}\sigma_z {A}_a^- \phi, 
\end{equation}
where 
\begin{equation}
{A}^-_{\mu}= - \frac{1}{2r(r+x^5)}{\eta}'_{\mu\nu i}x^{\nu}\tau^i,~~{A}^-_5=0, 
\end{equation}
with ${\eta}'_{\mu\nu i}=\epsilon_{\mu\nu i}-\eta_{\mu i}\eta_{\nu
4}+\eta_{\nu i}\eta_{\mu 4}$.  
The corresponding field strength is 
\begin{align}
&{F}^-_{\mu\nu}=\frac{1}{r^2}x_{\mu}{A}^-_{\nu}-\frac{1}{r^2}x_{\nu}{A}^-_{\mu}+\frac{1}{2r^2}{\eta}'_{\mu\nu i}\tau^i,\nonumber\\
&{F}^-_{\mu5}=-{F}^-_{5\mu}=-\frac{1}{r^2}(r+x^5){A}^+_{\mu}. 
\end{align}

With a different gauge choice $(x^5 < r)$, the holonomy is calculated as 
\begin{equation}
{A'}^+=-i\langle\!\langle{\chi'}_+|d\psi'_+\rangle
=-i\langle{\psi'_+}|d\chi' _+\rangle\!\rangle
=dx^a {\phi}^{\dagger}\sigma_z {A'}_a^+ \phi,  
\end{equation}
where  
\begin{equation}
{A'}^+_{\mu}=-\frac{1}{2r(r-x^5)}{\eta}'_{\mu\nu i}x^{\nu}\tau^i,~~{A'}^+_5=0. 
\end{equation}
The corresponding field strength is 
\begin{align}
&{F'}^+_{\mu\nu}=\frac{1}{r^2}x_{\mu}{A'}^+_{\nu}-\frac{1}{r^2}x_{\nu}{A'}^+_{\mu}-\frac{1}{2r^2}{\eta}'_{\mu\nu i}\tau^i,\nonumber\\
&{F'}^+_{\mu5}=-{F'}^+_{5\mu}=-\frac{1}{r^2}(r-x^5){A'}^+_{\mu}. 
\end{align}
Similarly, the lower energy degenerate states bring the holonomy  
\begin{equation}
{A'}^-=-i\langle\!\langle{\chi'}_-|d\psi'_-\rangle
=-i\langle{\psi'}_-|d\chi'_-\rangle\!\rangle
=dx^a {\phi}^{\dagger} \sigma_z {A'_a}^- {\phi},
\end{equation}
where 
\begin{equation}
{A_{\mu}'}^-=- \frac{1}{2r(r-x^5)} \eta_{\mu\nu i}x^{\nu}\tau^i,~~{A'_5}^-=0.
\end{equation}
The corresponding field strength is  
\begin{align}
&{F'_{\mu\nu}}^-=\frac{1}{r^2}x_{\mu} {A'}^-_{\nu}-\frac{1}{r^2}x_{\nu} {A'}^-_{\mu}+\frac{1}{2r^2}\eta_{\mu\nu i}\tau^i,\nonumber\\
&{F'_{\mu5}}^-=-{F'_{5\mu}}^-=\frac{1}{r^2}(r-x^5){A'}^+_{\mu}. 
\end{align}

The $SU(1,1)$ gauge transformation relates  $A^+$ and ${A'}^+$  as
\begin{equation}
\sigma_z {A'}^+_adx^a=g^{\dagger}_+(\sigma_z A^+_adx^a)  g_+-ig_+^{\dagger}\sigma_z dg_+, 
\end{equation}
and their field strengths as 
\begin{equation}
\sigma_z F'_{ab}=g^{\dagger}_+(\sigma_z F_{ab}) g_+, 
\end{equation}
where $g_{\pm}$ are given by (\ref{su11groupelements}). 
Similarly, $A^-$ and ${A'}^-$ are related by the $SU(1,1)$ transformation,  
\begin{equation}
\sigma_z {A'}^-_a dx^a =g_-^{\dagger}(\sigma_z A^-_a dx^a)  g_--ig_-^{\dagger}\sigma_z dg_-, 
\end{equation}
and 
\begin{equation}
\sigma_z {F'}^-_{ab}=g_-^{\dagger}(\sigma_z F^-_{ab}) g_-. 
\end{equation}
Thus, the exceptional points of the $SO(3,2)$ model act as the $SU(1,1)$ monopole with opposite charges for the upper and lower energy states.  
Such non-compact gauge group monopoles have been introduced in the context of the non-compact Hopf maps \cite{Hasebe2009}. 



\begin{thebibliography}{99}

\bibitem{Dyson1962}
Freeman J. Dyson,  
{\it``Statistical Theory of the Energy Levels of Complex Systems.''}, 
J. Math. Phys. 3 (1962) 140-156. 
\bibitem{Avron1988}
J. E. Avron, L. Sadun, J. Segert and B. Simon, 
{\it``Topological Invariants in Fermi Systems with TR Invariance''}, 
 Phys. Rev. Lett. 61 (1988) 1329-1332,   
{\it``Chern numbers, quaternions, and Berry's phases in Fermi systems''},  
Comm. Math. Phys.   
124 (1989) 595-627. 
\bibitem{TKNN1982} 
D. J. Thouless, M. Kohmoto, M. P. Nightingale, and M. den Nijs, 
{\it``Quantized Hall Conductance in a Two-Dimensional Periodic Potential''}, 
Phys. Rev. Lett. 49 (1982) 405-408.
\bibitem{Kohmoto1985}
 Mahito Kohmoto, 
{\it``Topological invariant and the quantization of the Hall conductance''}, 
Ann. Phys. 160 (1985) 343-354.
\bibitem{Bernevig2005}
B. Andrei Bernevig, Shou-Cheng Zhang 
{\it``Quantum Spin Hall Effect''}, Phys. Rev. Lett. 96 (2006) 106802;  cond-mat/0504147.   
\bibitem{Kane2005}
 C.L. Kane, E.J. Mele,   
{\it``$Z_2$ Topological Order and the Quantum Spin Hall Effect''},  
Phys. Rev. Lett. 95  (2005) 146802; cond-mat/0506581.  
\bibitem{FKM07}
L. Fu, C. L. Kane, and E. J. Mele,
{\it ``Topological Insulators in Three Dimensions''},
Phys. Rev. Lett. 98 (2007) 106803. 
\bibitem{MB07}
J. E. Moore and L. Balents
{\it ``Topological invariants of time-reversal-invariant band structures''},
Phys. Rev. B 75 (2007) 121306(R).
\bibitem{Roy09}
R. Roy, {\it ``Topological phases and the quantum spin Hall effect 
in three dimensions''},
Phys. Rev. B 79 (2009) 195322.
\bibitem{Qietal2008}
X-L. Qi, T. Hughes, S-C. Zhang, 
{\it``Topological Field Theory of TR Invariant Insulators''},  
Phys. Rev. B78 (2008) 195424-43; arXiv:0802.3537. 
\bibitem{Schnyderetal2008}
Andreas P. Schnyder, Shinsei Ryu, Akira Furusaki, Andreas W. W. Ludwig,  
{\it``Classification of topological insulators and superconductors in three spatial dimensions''},  
 Phys. Rev. B 78  (2008) 195125; arXiv:0803.2786.  
\bibitem{Kitaev2008}
A. Kitaev, 
{\it``Periodic table for topological insulators and superconductors''}, 
Proceedings of the L.D.Landau Memorial Conference Advances in Theoretical
Physics, June 22-26 (2008); arXiv:0901.2686.
\bibitem{FS97}
Y. V. Fyodorov and H.-J. Sommers,
{\it ``Statistics of resonance poles, phase shifts and time delays in
	quantum chaotic scattering: Random matrix approach for systems
	with broken time-reversal invariance''},
J. Math. Phys. 38 (1997) 1918-1981.
\bibitem{GT85}
R. Graham and T. T\'{e}l,
{\it ``Quantization of H\'{e}nons map with dissipation''},
Z. Phys. B 60 (1985) 127-136.
\bibitem{HatanoNelson199678}
 Naomichi Hatano, David R. Nelson, 
 {\it``Localization transitions in non-hermitian quantum mechanics''},  
Phys. Rev. Lett. 77  (1996) 570-573; cond-mat/9603165,  
{\it``Vortex Pinning and non-hermitian Quantum Mechanics''},  
 Phys. Rev. B 56  (1997) 8651-8673; cond-mat/9705290,   
{\it``non-hermitian Delocalization and Eigenfunctions''},  
 Phys. Rev. B 58 (1998) 8384-8390; cond-mat/9805195.  
\bibitem{BenderBoettcher1998} 
  Carl M. Bender, Stefan Boettcher
 {\it``Real Spectra in non-hermitian Hamiltonians Having PT Symmetry''}, 
Phys. Rev. Lett. 80 (1998) 5243-5246; physics/9712001. 
 Carl Bender, Stefan Boettcher, Peter Meisinger,  
{\it``PT-Symmetric Quantum Mechanics''},  
 J. Math. Phys. 40 (1999) 2201-2229; quant-ph/9809072. 
Carl M Bender, 
{\it``Making sense of non-hermitian Hamiltonians''},  
Rep. Prog. Phys. 70 (2007) 947. 
\bibitem{Mostafazadeh2008}
See as a review, 
 Ali Mostafazadeh,  
{\it``pseudo-hermitian Representation of Quantum Mechanics''},  
  Int. J. Geom. Meth. Mod. Phys 7 (2010) 1191-1306; arXiv:0810.5643.  
\bibitem{BG11}
D. C. Brody and E.-M. Graefe, 
{\it ``On complexified mechanics and coquaternions''}
J. Phys. A {\bf 44}, 072001 (2011).
\bibitem{BG11_3}
D. C. Brody and E.-M. Graefe, 
{\it ``Coquaternionic quantum dynamics for two-level systems''}
arXiv:1105.4038, to be published in Acta Polytechnica.
\bibitem{Cockle1848}
 James Cockle,
 {\it``On Certain Functions Resembling Quaternions, and on a New Imaginary Algebra''},
 Phil. Mag. (3) 33 (1848) 435-439, 
{\it``On a New Imaginary in Algebra''}, Phil. Mag. (3) 34 (1849) 37-47, 
{\it``On Systems of Algebra Involving more than one Imaginary and on Equations of the Fifth Degree''}, Phil. Mag. (3) 35 (1849) 434-437.
\bibitem{Hamilton1844}
William Rowan Hamilton, 
{\it``On a new Species of Imaginary Quantities connected with a theory of Quaternions''}, 
Proceedings of the Royal Irish Academy, 2 (1844), pp. 424-434.
\bibitem{FaisalandMoloney1981}
F H M Faisal and J V Moloney, 
{\it``Time-dependent theory of non-hermitian Schrodinger equation: Application to multiphoton-induced ionisation decay of atoms''},  
 J. Phys. B: At. Mol. Phys. 14 (1981) 3603. 
\bibitem{Wong1967}
Jack Wong, 
{\it``Results on Certain non-hermitian Hamiltonians''}, 
J. Math. Phys. 8  (1967) 2039. 
\bibitem{MehtaBook}
M. Mehta, {\it``Random Matrices''}, Third Edition 2004, Elsevier.
\bibitem{Zirnbauer1996}
M. Zirnbauer, 
{\it``Riemannian symmetric superspaces and their origin in random-matrix theory''}, 
J. Math. Phys. 37 (1996), 4986-5018.
\bibitem{AtlandZirnbauer1997}  
 Alexander Altland and Martin R. Zirnbauer,  
{\it``Nonstandard symmetry classes in mesoscopic normal-superconducting hybrid structures''}, 
Phys. Rev. B 55  (1997)1142-1161.
\bibitem{Bernard2001}
D. Bernard and A. LeClair, {\it``A classification of 
non-hermitian random matrices''}, arXiv:cond-mat/0110649.
\bibitem{BenAryeh2004} 
Yacob Ben-Aryeh, Ady Mann and Itamar Yaakov, 
{\it``Rabi oscillations in a two-level atomic system with a pseudo-hermitian Hamiltonian''}, J. Phys. A: Math. Gen. 37 (2004) 12059-12066.  
\bibitem{Mason2006}
 Lionel J. Mason 
{\it``Global anti-self-dual Yang-Mills fields in split signature and their scattering''}, J. Reine Angew. Math. 597 (2006) 105-133, math-ph/0505039.
\bibitem{HasebeTwistor2010}
Kazuki Hasebe,
{\it``Split-Quaternionic Hopf Map, Quantum Hall Effect, and Twistor Theory''}, 
Phys. Rev. D 81 (2010) 041702(R);  arXiv:0902.2523. 
\bibitem{BBJ2002}
C. M. Bender, D.C.Brody, and H. F. Jones,
{\it ``Complex extension of quantum mechanics" }
Phys. Rev. Lett., 89 (2002) 270401.
\bibitem{Ahmed2003}
Zafar Ahmed,  
{\it``C-, PT- and CPT-invariance of pseudo-hermitian Hamiltonians''},  
 J. Phys. A: Math. Gen. 36 (2003) 9711; quant-ph/0302141. 
\bibitem{Berry1984}
M. V. Berry,  
{\it``Quantal Phase Factors Accompanying Adiabatic Changes''}, 
Proc. R. Soc. Lond. A 392 (1984) 45-57.
\bibitem{Nesterov2008} 
 Alexander I. Nesterov, F. Aceves de la Cruz,  
{\it``Complex magnetic monopoles, geometric phases and quantum evolution in vicinity of diabolic and exceptional points''},  
  J. Phys. A41 (2008) 485304; arXiv:0806.3720. 
\bibitem{Hasebe2009} 
 Kazuki Hasebe,  
{\it``The Split-Algebras and Non-compact Hopf Maps''}, 
 J. Math. Phys. 51 (2010) 053524; arXiv:0905.2792.  
\bibitem{Wilczek1984}
Frank Wilczek and A. Zee,
{\it``Appearance of Gauge Structure in Simple Dynamical Systems''}, 
Phys. Rev. Lett. 52  (1984) 2111-2114. 
\bibitem{Luttinger}
J. M. Luttinger, 
{``\it Quantum Theory of Cyclotron Resonance in Semiconductors:
General Theory''}, Phys. Rev. 102, 1030 (1956).
\bibitem{Murakamietal2003}
Shuichi Murakami, Naoto Nagaosa, Shou-Cheng Zhang,  
{\it``Dissipationless Quantum Spin Current at Room Temperature''},  Science 301, 1348 (2003),  {\it``SU(2) Non-Abelian Holonomy and Dissipationless Spin Current in Semiconductors''},  
  Phys. Rev. B69, 235206 (2004); cond-mat/0310005.
\end{thebibliography}
\end{document}